\documentclass{JHEP3}
\usepackage{graphicx}
\usepackage{amssymb}
\usepackage{amsmath}
\usepackage{cite}
\usepackage{afterpage}

\newcommand{\lta}{\lesssim}
\newcommand{\gta}{\gtrsim}

\newcommand{\be}{\begin{equation}}
\newcommand{\en}{\end{equation}}
\newcommand{\bea}{\begin{eqnarray}}
\newcommand{\ena}{\end{eqnarray}}

\newcommand{\mP}{m_{_{\mathrm{Pl}}}}

\newcommand{\mean}[1]{\left\langle #1 \right\rangle}
\newcommand{\mpl}{m_{_\mathrm{Pl}}}

\newcommand{\ie}{{\sl i.e.}~}

\title{The SUGRA Quintessence Model Coupled to the MSSM}

\author{Philippe Brax \thanks{Associate Researcher at Institut
d'Astrophysique de Paris, UMR 7095-CNRS, Universit\'e Pierre et Marie
Curie, 98bis boulevard Arago, 75014 Paris, France} \\ Service de
Physique Th\'eorique, CEA-Saclay, Gif/Yvette cedex, France F-91191 \\
E-mail: \email{brax@spht.saclay.cea.fr}}

\author{J\'er\^ome Martin \\ Institut d'Astrophysique de Paris, UMR
7095-CNRS, Universit\'e Pierre et Marie Curie, 98bis boulevard Arago,
75014 Paris, France \\ E-mail: \email{jmartin@iap.fr}} \date{\today}

\abstract{We study the cosmological evolution of the universe when
quintessence is modeled within supergravity, supersymmetry is broken
in a hidden sector, and we also include observable matter in a third
independent sector. We find that the presence of hidden sector
supersymmetry breaking leads to modifications of the quintessence
potential. We focus on the coupling of the SUGRA quintessence model to
the MSSM and investigate two possibilities.  First one can preserve
the form of the SUGRA potential provided the hidden sector dynamics is
tuned. The currently available limits on the violations of the
equivalence principle imply a universal bound on the vacuum
expectation value of the quintessence field now, $\kappa ^{1/2}Q\ll
1$. On the other hand, the hidden sector fields may be stabilised
leading to a minimum of the quintessence potential where the
quintessence field acquires a mass of the order of the gravitino mass,
large enough to circumvent possible gravitational problems. However,
the cosmological evolution of the quintessence field is affected by
the presence of the minimum of the potential. The quintessence field
settles down at the bottom of the potential very early in the history
of the universe. Both at the background and the perturbation levels,
the subsequent effect of the quintessence field is undistinguishable
from a pure cosmological constant.}

\begin{document}

\section{Introduction}

There is a host of observational evidence in favor of the existence of
a non-zero vacuum energy density of the Universe driving the
acceleration of the expansion of the Universe~\cite{LSS,IA,CMB}. The
simplest explanation for this new era in the history of the Universe
is the presence of a cosmological constant of extraordinarily small
value, some 120 orders of magnitude lower than the Planck scale. Such
a small value is particularly difficult to accommodate when dynamical
effects such as the Quantum Chromo-Dynamics (QCD) and electroweak
phase transitions or even Grand Unified Theory (GUT) scale physics are
taken into account. This has prompted the possibility of using extra
dimensional models such as self--tuning scenarios~\cite{kachru} or
brane induced gravity models~\cite{deffayet}. Unfortunately these
alternatives have drawbacks such as hidden
fine--tunings~\cite{Lalak}. Within four dimensional physics, there is
an experimental way of discovering whether the vacuum energy is a true
constant of nature or the result of more complicated dynamical
effects. Indeed very active experimental programs are dedicated to the
analysis of the so-called equation of state of the dark energy sector
(the ratio between the pressure and the energy density). If the
equation of state differs from $-1$ (and is greater than $-1$,
otherwise see for instance Ref.~\cite{MSU}), then a plausible
candidate for dark energy is quintessence~\cite{RP,quint,PB}, \ie the
dynamics of a scalar field rolling down a runaway potential. Of
course, quintessence only accounts for the small and non-vanishing
vacuum energy, it has nothing to say about the large cancellation of
the overall cosmological constant.

\par

One of the most stringent requirements imposed on quintessence models
is the existence of attractors, \ie long time stable solutions of the
equations of motion~\cite{quint}. Indeed the presence of an attractor
implies an insensitivity to initial conditions of the quintessence
field for the vacuum energy now.  For a large class of potentials,
attractors leading to vacuum energy dominance exist provided their
large field behavior is of the inverse power law type. Such potentials
are known under the name of Ratra--Peebles potentials~\cite{RP}. In
these cases, the quintessence field reaches an attractor, only to
leave it when dominating the energy content of the universe. This
happens when the field is of the order of the Planck scale.

\par

This has drastic consequences on quintessence model building. Indeed,
it requires a natural framework within which Planck scale physics is
taking into account. Supergravity is a promising field theoretical
arena where both particle physics and Planck scale physics can be
described~\cite{Nilles}. Models of quintessence in supergravity have
been constructed~\cite{BM1,BM2,BMR1,BMR2} leading to interesting
phenomenological consequences such as low values of the equation of
state. In particular, the simplest model of quintessence in
supergravity, often dubbed the SUGRA model in the literature, leads to
the following potential
\begin{equation}
\label{sugra}
V_{\rm quint}(Q)={\rm e}^{\kappa Q^2/2+\kappa \xi
^2}\frac{M^{4+\alpha }}{Q^{\alpha}}\, ,
\end{equation}
with $\kappa \equiv 8\pi/\mpl ^2$ and $M^{4+\alpha }=\lambda ^2\xi ^4
m_{\rm c}^{\alpha }2^{\alpha/2}$ and where, in this equation, $Q$ is
canonically normalized. The quantity $\alpha $ is a free positive
index and $\lambda $ is a dimensionless coupling constant and, in
order to avoid any fine-tuning, we will always consider that $\lambda
\sim 1$. $m_{\rm c}$ is the cut-off scale of the effective theory used
in order to derive the SUGRA potential. Typically $m_{\rm c}$ can be
thought as the GUT scale but we will see that the Planck scale is also
possible (and, sometimes, necessary). Finally, $\xi $ is a vacuum
expectation value (vev) of some other field present in the
quintessence sector, see below for more details. As a specific
example, $\xi$ can be realized as a Fayet-Iloupoulous term arising
from the Green--Schwarz anomaly cancellation
mechanism~\cite{BM1,BM2}. The main feature of the above potential is
that supergravity corrections have been exponentiated and appear in
the prefactor. Phenomenologically, this potential has the nice feature
that the equation of state $\omega \equiv p_Q/\rho _Q$ can be closer
to $-1$ than with the Ratra--Peebles potential when the field
approaches its present value $\kappa^{1/2} Q_{\rm now}\approx
1$. Moreover, a small value for $M$ can be avoided. Indeed, since the
vev of the quintessence field is now of the order of the Planck mass,
requiring that the quintessence energy density be of the order of the
critical energy density $\rho_{\rm cri}\sim 10^{-122}\mpl ^4$ implies
that
\begin{equation}
\frac{M}{\mpl}\sim 10^{-122/(4+\alpha )}\, ,
\end{equation}
and, therefore, $M$ can be a large scale (by particle physics
standard) for very reasonable values of the index $\alpha $. For
instance, it is above the TeV scale for $\alpha \gta 4$. This
mechanism is reminiscent of a ``see-saw'' mechanism where a very small
scale (the cosmological constant scale) is explained in terms of a
large one (the scale $M$) and a very large one (the Planck scale
$m_{_{\rm Pl}}$).  Moreover, the scale $\xi $ can have acceptable
values. From the expression of the scale $M$, one obtains
\begin{equation}
\label{xisugra}
\frac{\xi }{\mpl }\sim \left(\frac{\rho _{\rm cri}}{\mpl
}\right)^{1/4}
\left(\frac{\mpl}{m_{\rm c}}\right)^{\alpha /4}\, ,
\end{equation}
and for $\alpha \gta 11$ and a cut-off $m_{\rm c}$ of the order of the
GUT scale, $\xi $ is above the TeV scale\cite{BM1}. However,
considering $m_{\rm c}<m_{_{\rm Pl}}$ can also be viewed as
problematic since, as already mentioned, the vev of the quintessence
field tends to be of the order of the Planck mass. In this case, it is
difficult to control the shape of the K\"ahler potential, see
Eq.~(\ref{quintsector}). Facing this issue, a natural choice could be
$m_{\rm c}=m_{_{\rm Pl}}$. Then the above equation indicates that $\xi
$ needs to be fine-tuned as the same level as the cosmological
constant even if the model remains different since the equation of
state is redshift dependent. Moreover, in this situation, the choice
of the K\"ahler potential strongly influences the shape of the scalar
potential~\cite{BM2} since the SUGRA corrections are of order
one. Hence, it is no longer possible to see the K\"ahler potential of
Eq.~(\ref{quintsector}) as a Taylor expansion but it should rather be
considered as a specific choice. Another route is to argue that, in
these circumstances, the quintessence energy density can easily remain
less than $m_{\rm c}^4$, indicating that the theory can still be
meaningful if the cut-off is interpreted as the cut-off to the energy
density scales and not to the vev's of the fields. In any case, it is
clear that there are some fine-tuning problems at this level even if it
is arguably less acute than in the cosmological constant case.

\par

However, it is clear that the quintessence sector cannot be considered
as disconnected from the particle physics standard model (or its
extensions) and should be embedded in a more general structure. In
Ref.~\cite{BMpart}, we have investigated the coupling between the
quintessence sector and a hidden sector where supersymmetry is
broken. A general formalism to calculate the corresponding
implications was presented. On very general grounds, it was shown
that, as a consequence of this coupling, the shape of the quintessence
potential is changed and that the particle masses become dependent on
the quintessence field which implies the presence of a fifth force, a
violation of the equivalence principle and possibly, depending on the
complexity of the model, a variation of the gauge couplings and of the
proton to electron mass ratio.

\par

The main goal of the present article is to apply this general
formalism to a concrete case, namely the SUGRA one, where the
quintessence potential is described by Eq.~(\ref{sugra}). We find that
the coupling between the quintessence, observable and hidden sectors
has tremendous consequences on the dynamics of the quintessence
field. The first one is that the shape of the quintessence potential
is modified.  When the hidden sector fields have been stabilized, the
potential acquires a minimum located at a very small value of the
field (in comparison with the Planck mass). As a result, the model
becomes equivalent to a pure cosmological constant as the quintessence
field settles at the minimum of a potential before Big Bang
Nucleosynthesis (BBN). The mass of the field is also changed and
becomes equal to the gravitino mass. The above conclusion is also true
at the perturbative level as no growing mode is present despite the
smallness of the Jeans length.  On top of this, the energy scales in
the quintessence sector have to be fine--tuned at a highest level than
the cosmological constant itself. This makes the whole scenario very
unappealing.  On the other hand, preserving the shape of the SUGRA
potential requires a fine-tuning of the hidden sector dynamics. If
this is done, one may wonder whether the model is still
phenomenologically acceptable. We show that, in this case, the
scenario tends to be ruled out by local tests of gravity rather than
by cosmological considerations.

\par

Our results have been obtained using the SUGRA model~\cite{BM1} of
quintessence. Effectively the only crucial ingredient is the fact that
the potential in the quintessence sector reduces to the Ratra--Peebles
form for small values of the quintessence field.  Within this
framework, \ie a quintessence sector with a Ratra--Peebles potential
at small values of the quintessence field coupled to a hidden sector
gravitationally, our conclusions apply and are generic even if a
complete scan of the parameter space $(m_{3/2},m_{1/2}, \omega _Q)$
has not yet been performed.

\par

The paper is arranged as follows. In the following section, \ie
section~\ref{Quintessence and Supersymmetry Breaking}, we discuss the
coupling of SUGRA quintessence to the hidden sector of supersymmetry
breaking and the observable sector. In particular we find that the
generic quintessence potential has a minimum with a mass for the
quintessence field of the order of the gravitino mass.  On the other
hand one can fine--tune the hidden sector dynamics to preserve the
form of the SUGRA potential.  In section~\ref{Implications for Gravity
Experiments}, we investigate the implications for gravity experiments
of this fine--tuning of the SUGRA potential. In
section~\ref{Implications for Cosmology}, we discuss the cosmological
evolution of the generic case where the quintessence potential
develops a minimum and show that the field settles at the minimum of
the potential before nucleosynthesis. We then examine the cosmological
perturbations around the minimum of the potential and show that there
are no growing modes for quintessence perturbations. We then conclude
and mention possible way-outs in section~\ref{Conclusions}.

\section{Quintessence and Supersymmetry Breaking}
\label{Quintessence and Supersymmetry Breaking}

\subsection{The Framework}
\label{The Framework}

As described in the Introduction, the usual approach consists in
taking into account the features of the quintessence sector
only. However, the non-observation of supersymmetric partners to the
standard model particles implies that SUSY must be broken at a scale
of the order of a TeV. One must also take into account the existence
of an observable sector, modeled as the Minimal Super Symmetric Model
(MSSM) or mSUGRA model~\cite{Nilles}, where the standard model
particles live. The calculations based on this more realistic
description are performed in Ref.~\cite{BMpart} where the model is
separated in three sectors. There is a quintessence sector as already
presented. It couples gravitationally to a visible sector where the
standard model particles live. The supersymmetry breaking sector is
also interacting gravitationally with the other two sectors. At the
level of the K\"ahler $K$ and super potentials $W$, it is a simple sum
of the contributions from each sector.
\begin{equation}
K= K_{\rm quint} + K_{\rm hid}+ K_{\rm obs}\, , \quad W= W_{\rm quint}
+ W_{\rm hid} + W_{\rm obs}\, .
\end{equation}
In the present article, we use the general results obtained in
Ref.~\cite{BMpart} and apply them to a specific class of quintessence
model.

\par

The quintessence sector is chosen such that it gives the SUGRA version
of the Ratra-Peebles potential. It has been shown in Ref.~\cite{BM1}
that this can be obtained using
\begin{eqnarray}
\label{quintsector}
K_{\rm quint} &=& QQ^{\dagger} +XX^{\dagger} +
YY^{\dagger}\frac{\left(QQ^{\dagger }\right)^p}{m_{\rm c}^{2p}}+\sum
_{\alpha =1}^n\left(X_{\alpha }X_{\alpha }^{\dagger }+Y_{\alpha
}Y_{\alpha }^{\dagger }\right)+\cdots \, ,\\ W_{\rm quint} &=& \lambda
X^2Y+\sum _{\alpha =1}^n\lambda _{\alpha }X_{\alpha }^2Y_{\alpha
}+\cdots \, ,
\end{eqnarray}
where $X_{\alpha }$ and $Y$, $Y_{\alpha }$ are superfields satisfying
$\mean{X}=\xi$, $\mean{X_{\alpha }}=\xi _{\alpha}$ and
$\mean{Y}=\mean{Y_{\alpha }}=0$ at the GUT scale where the model is
defined (of course, in this context, the GUT scale is just an
illustration), see however the discussion after
Eq.~(\ref{xisugra}). This implies that $\mean{W_{\rm quint}}=0$ and
guarantees the positivity of the potential in the quintessence
sector. The quantities $\lambda$, $\lambda _{\alpha }$ are
dimensionless coupling constants and $p$ is a free index. As already
mentioned, the scale $m_{\rm c}$ is the GUT scale below which the
theory under consideration is valid. This means that the theory is
valid only for vevs that are much less than $m_{\rm c}$ or energy
densities less than $m_{\rm c}^4$ according to the interpretation
given to the cut-off. The influence of the K\"ahlerian corrections are
a priori important and have been studied in Ref.~\cite{BM2}. In the
above expression and in the rest of this paper the dots stand for the
higher order terms suppressed by $m_{\rm c}$. In the ``dark energy''
sector, we collectively denote the fields by $d_{\alpha
}=X,Y,X_{\alpha },Y_{\alpha },Q$. Let us notice that we have assumed
that the K\"ahler potential in the quintessence sector is regular at
the origin. This excludes the no-scale case which deserves a special
treatment, see Ref.~\cite{BMnoscale}. It is interesting to consider
regular K\"ahler potentials since they naturally lead to inverse power
law scalar potentials while no scale K\"ahler potentials tend to give
exponential potentials with very different properties, see
Ref.~\cite{BMnoscale}.

\par

For the hidden sector, we follow Ref.~\cite{BMpart}. We denote the
fields in the hidden sector by $z_i$ and assume that $K_{\rm hid}$ is
regular for small values of the hidden fields and can be Taylor
expanded. Without specifying the superpotential for the moment, this
leads to
\begin{equation}
K_{\rm hid}=\sum _i z_iz_i^{\dagger }+\cdots \, ,\quad W_{\rm
  hid}=W_{\rm hid}(z_i)\, ,
\end{equation}
where, as before, the dots denote the higher order terms that are not
considered in this article.
\par

Finally, the fields in the matter sector are written $\phi _a$.  This
sector is supposed to contain all the (super) fields that are
observable (including the dark matter). As a consequence, following
again Ref.~\cite{BMpart}, we take this sector to be the MSSM or the
mSUGRA model~\cite{Nilles}, that is to say
\begin{equation}
\label{wobs}
K_{\rm obs}=\sum _a\phi _a\phi _a^{\dagger }+\cdots \, ,\quad W_{\rm
obs}=\frac13 \sum _{abc}\lambda _{abc}\phi_a \phi_b \phi_c +\frac12
\sum _{ab} \mu _{ab}\phi _a\phi _b\, ,
\end{equation}
with a supersymmetric mass matrix $\mu_{ab}$ and Yukawa couplings
$\lambda_{abc}$. In order to completely specify the observable sector,
it is also necessary to choose the supergravity gauge coupling
functions $f_{_{G}}$. {\it A priori}, all the $f_{_{G}}$'s are $z_i$
and $d_{\rm \alpha }$--dependent. If, indeed, these functions are not
constant, then this implies variations of the coupling constants. In
particular, this leads to a variation of the fine structure constant,
see Ref.~\cite{BMpart}.

\par

Let us now discuss the breaking of supersymmetry in more details. In
the hidden sector, the supersymmetry breaking fields $z_i$ take a vev
determined by
\begin{equation}
\partial_{z_i} V=0
\end{equation}
where $V$ is the total potential obtained from the previous model. The
presence of the quintessence field affects the dynamics of the hidden
sector and the vev's of the hidden sector fields become a priori $Q$
dependent. They are parameterized in a model independent way by the
coefficients $a_i(Q)$ and $c_i(Q)$ according to
\begin{equation}
\label{parahidden}
\kappa ^{1/2}\mean{z_i}_{\rm min}\sim a_i(Q)\, , \quad \kappa
    \mean{W_{\rm hid}}_{\rm min}\sim M_{_{\rm S}}(Q)\, , \quad \kappa
    ^{1/2}\mean{\frac{\partial W_{\rm hid}}{\partial z_i}}_{\rm
    min}\sim c_i(Q)M_{_{\rm S}}(Q)\, ,
\end{equation}
where $M_{_{\rm S}}$ is the supersymmetry scale. Notice that, if the
cut-off of the theory is much less than $m_{_{\rm Pl}}$, then we
expect $a_i\ll 1$ since the vev's of $\langle z_i\rangle $ must be at
most of the order of $m_{\rm c}$. Notice also that, a priori, nothing
can be said about the function $c_i(Q)$. On the other hand, if the
cut-off is close to the Planck mass, the dynamics of the hidden sector
cannot be analyzed in a model independent way and we must rely on
particular models for the hidden sector to go any further. This is not
surprising (and is even to be expected) since the situation is in fact
exactly similar to what happens in the standard case of the mSUGRA
model where the hidden sector is often described by the Polonyi model
and the hidden field is stabilized at a vev of the order of the Planck
mass. However, in the standard case, our ignorance of the hidden
sector is parameterized in terms of two numbers, $a_i$ and $c_i$, or
equivalently $m_{3/2}$ and $m_{1/2}$. In presence of dark energy, the
situation is more complicated since our ignorance of the hidden sector
is now described by two free functions $a_i(Q)$ and $c_i(Q)$.

\par

Another quantity of interest is of course the gravitino mass. It is
defined by the following expression
\begin{equation}
m_{3/2}\equiv \left\langle \kappa W {\rm e}^{\kappa K/2}\right\rangle
  \, ,
\end{equation}
where $K$ and $W$ are the total K\"ahler and super potentials (\ie
taking into account the three sectors). In the present context, the
gravitino mass may depend on the vev of the quintessence
field. Therefore, it is natural to write
\begin{equation}
\label{gravitino}
m_{3/2}={\rm e}^{\kappa K_{\rm quint}/2+\sum _i\vert a_i\vert
^2/2}M_{_{\rm S}}\equiv {\rm e}^{\kappa K_{\rm quint}/2}m_{3/2}^0\, ,
\end{equation}
where $m_{3/2}^0$ is the mass that the gravitino would have without
the presence of the quintessence field. This quantity explicitly
appears in the expression of the scalar potential. Let us notice that
$W_{\rm quint}$ does no appear in the previous formula because one has
$\mean{W_{\rm quint}}=0$ for the particular case of the SUGRA model
considered here. 

\subsection{The Dark Energy Sector}
\label{The Dark Sector}

In this subsection, we study how the shape of the SUGRA model is
changed by supersymmetry breaking. As discussed in Ref.~\cite{BMpart},
see Eq.~(2.18) of that article, the new potential in the dark sector
is given by
\begin{eqnarray}
\label{potde}
V_{_{\rm DE}} &=& {\rm e}^{\sum _i\vert a_i\vert ^2} V_{\rm quint} +
M_{_{\rm S}}^2 {\rm e}^{\kappa K_{\rm quint}+\sum _i\vert a_i\vert ^2}
\biggl[\left(K^{-1}\right)^{d_{\alpha }^{\dagger}d_{\beta }}
\frac{{\partial }K_{\rm quint}}{\partial d_{\beta }} \frac{{\partial
}K_{\rm quint}}{\partial d_{\alpha }^{\dagger }} -\frac{3}{\kappa
}\biggr] \nonumber \\ & & + M_{_{\rm S}} {\rm e}^{\kappa K_{\rm
quint}+\sum _i\vert a_i\vert ^2}
\Biggl\{\biggl[\left(K^{-1}\right)^{d_{\alpha }^{\dagger}d_{\beta }}
\frac{{\partial }K_{\rm quint}}{\partial d_{\beta }} \frac{{\partial
}K_{\rm quint}}{\partial d_{\alpha }^{\dagger }}-\frac{3}{\kappa
}\biggr] \biggl(\kappa W_{\rm quint} +\kappa W_{\rm quint}^{\dagger
}\biggr) \nonumber \\ & & + \left(K^{-1}\right)^{d_{\alpha
}^{\dagger}d_{\beta }}\biggl( \frac{{\partial }K_{\rm quint}}{\partial
d_{\beta }} \frac{{\partial }W_{\rm quint}^{\dagger }}{\partial
d_{\alpha }^{\dagger }} + \frac{{\partial }K_{\rm quint}}{\partial
d_{\alpha }^{\dagger }} \frac{{\partial }W_{\rm quint}}{\partial
d_{\beta }}\biggr)\Biggr\}+\sum _i\vert F_{z_i}\vert ^2\, ,
\end{eqnarray}
where $V_{\rm quint}$ is the potential that one would have obtained by
considering the dark energy sector alone and $F_{z_i}$ the F--term in
the hidden sector which takes the form
\begin{eqnarray}
\label{defFz}
F_{z_i} &=& {\rm e}^{\kappa K_{\rm quint}/2+\sum _i\vert a_i\vert
^2/2} \frac{1}{\kappa ^{1/2}}\biggl[ \left(M_{_{\rm S}}+\kappa \left
\langle W_{\rm quint}\right \rangle \right)a_i+M_{_{\rm
S}}c_i\biggr]\, .
\end{eqnarray}
In the case of the model described by Eqs.~(\ref{quintsector}), the
above expressions simplify a lot, as $\mean{W_{\rm quint}}=0$.
Instead of the usual shape of the SUGRA potential, $V(Q)={\rm
e}^{\kappa \left(Q^2+\xi^2+\sum _{\alpha }\xi _{\alpha }^2\right)}\\
\times M^{4+2p}Q^{-2p}$, see Eq.~(\ref{sugra}) (in the minimal
approach, the fields $X_{\alpha }$ are not present which explains the
absence of the term $\sum _{\alpha }\xi _{\alpha }^2$ in the above
mentioned formula), we now have
\begin{equation}
\label{newpotde}
V_{_{\rm DE}}(Q)={\rm e}^{\kappa \left( Q^2+\xi ^2+\sum _{\alpha }\xi
  _{\alpha
  }^2\right)}\left[\frac{M^{4+2p}}{Q^{2p}}+\left(m_{3/2}^0\right)^2Q^2-\Upsilon
  ^4(Q)\right]\, ,
\end{equation}
where the function $\Upsilon (Q)$ encodes our ignorance of the hidden
sector and is given by 
\begin{eqnarray}
\label{defLamb} -\Upsilon ^4 (Q)&=& \left(m_{3/2}^0\right)^2
\left(\xi ^2+\sum _{\alpha =1}^n\xi _{\alpha }^2\right)+{\rm e}^{\sum
  _i\vert a_i\vert ^2}\sum _{\alpha =1}^n\lambda _{\alpha}^2\xi
  _{\alpha }^4 \nonumber \\ & & +\frac{1}{\kappa
  }\left(m_{3/2}^0\right)^2\left\{\sum
  _i\left[a_i(Q)+c_i(Q)\right]\right\}^2 -\frac{3}{\kappa
  }\left(m_{3/2}^0\right)^2\, ,
\end{eqnarray}
and $M^{4+2p}={\rm e}^{\sum _i\vert a_i\vert ^2}\lambda ^2\xi ^4
m_{\rm c}^{2p}$. In fact the quintessence field is not correctly
normalized in this expression.  To obtain a correctly normalized
field, it is sufficient to replace $Q$ with $Q/\sqrt{2}$. The new
shape of the potential is still not fixed in the above expression and
is only known when the functions $a_i(Q)$ and $c_i(Q)$ are specified,
\ie when the hidden sector is known explicitly.

\par

We can envisage two different situations. They are distinguished by
the equation of state $w_Q$ of the quintessence sector when $Q$ takes
its present value in the history of the universe. First of all, let us
assume that the equation of state is $w_Q\neq -1$. This can only be
achieved when the potential is of runaway type with an effective mass
for the quintessence field $m_Q\sim H_0$ the present Hubble rate. This
situation can only be achieved when the functions $a_i(Q)$ and
$c_i(Q)$ are not constant and such that, despite the new terms which
modify the shape of $V_{_{\rm DE}}$, the runaway shape of the
potential is preserved. This means that the susy breaking fields $z_i$
are not stabilized or, more precisely, that the fields follow a
trajectory in the field space $(z_i,Q)$. In this case, as we discuss
in the following, one should precisely evaluate how serious the
gravitational problems are. Obviously, this cannot be done in detail
unless the functions $a_i(Q)$ and $c_i(Q)$ are known exactly. In the
following we will emphasize the case where $a_i\ll 1$ and $c_i$ is
tuned to obtain a runaway potential.

\par

Another case of particular interest is when a satisfactory model of
the hidden sector has been found and the $z_i$'s are correctly
stabilized, \ie $a_i(Q)$ and $c_i(Q)$ are independent of $Q$. In
addition, if we assume that the fields in the hidden sector are
stabilized at vev's compatible with the cut-off $m_{\rm c}$ of the
theory, \ie we assume that $\mean{z_i}\ll m_{\rm c}$, and if $m_{\rm
c}\ll m_{_{\rm Pl}}$, then the coefficients $a_i$ are very small
(since $m_{\rm c}\ll \mpl$) but this does not imply anything about the
coefficients $c_i$. On the other hand, if $m_{\rm c}$ is not small in
comparison to $m_{_{\rm Pl}}$, then nothing can be said about $a_i$.

\par

Let us now focus on the simplest model of supersymmetry breaking where
there is only one field $z$ with a flat K\"ahler potential and a
constant superpotential, $W_{\rm hid}=m^3$\cite{Ibanez}. Then, one can
even justify the previous choice. Indeed, first of all, this
immediately implies that $M_{_{\rm S}}=\kappa m^3$ and $c=0$. Then,
the total potential (for simplicity, here, we assume $\lambda _{\alpha
}=0$) reads
\begin{eqnarray}
V &=&{\rm e}^{\kappa K_{\rm quint}}{\rm e}^{\kappa zz^{\dagger }}
\kappa m^6\left[\kappa zz^{\dagger }-3+\kappa \left(\xi^2+\sum
_{\alpha =1}^n\xi_{\alpha }^2+QQ^{\dagger }\right)+\frac{\lambda ^2
\xi ^4m_{\rm c}^{2p}}{\kappa m^6(Q Q^\dagger)^p}+\cdots \right]\, ,
\nonumber \\
\end{eqnarray}
where the dots now indicate the part containing the observable sector
terms. These terms do not play a role at high energy and therefore can
be ignored in the present context. At the minimum, we have $\partial
V/\partial z^{\dagger }=0$, that is to say
\begin{equation}
{\rm e}^{\kappa K_{\rm quint}}{\rm e}^{\kappa zz^{\dagger }} \kappa ^2
m^6 z\left[\kappa zz^{\dagger }-2+\kappa \left(\xi^2+\sum _{\alpha
=1}^n\xi_{\alpha }^2+QQ^{\dagger }\right)+\frac{\lambda ^2 \xi
^4m_{\rm c}^{2p}}{\kappa m^6(Q Q^\dagger)^p}\right]=0\, .
\end{equation}
The constraint coming from the smallness of the vacuum energy implies
(this is just another manifestation of the fact that quintessence has
nothing to say about the cosmological constant problem)
\begin{equation}
\label{sumxial} \xi^2 + \sum _{\alpha =1}^n\xi _{\alpha }^2\sim
\frac{3}{\kappa }\, ,
\end{equation}
and, therefore, the quantity in the squared bracket is positive. As a
result, the only solution is
\begin{equation}
\mean{z}_{\rm min}=0\, ,
\end{equation}
\ie $a=0$. Hence, we find that this simple model satisfies $a_i=c_i=0$
and, moreover, gives a SUSY breaking scale that does not depend on the
quintessence field. In this case the function $\Upsilon (Q)$ becomes a
constant and the potential in $Q$ admits a minimum. At this minimum,
it is clear that the mass of the quintessence field is of the order of
the gravitino mass, $m_Q\sim m_{3/2}= O(1) \mbox{TeV}$. In more
complex settings, one can envisage a case where one of the previous
results is relaxed, for instance where $c(Q)$ becomes $Q$-dependent
while the other functions describing the hidden sector remain
constant. This is what will be done in the following.

\par

We have thus reached one important conclusion, namely that when the
quintessence sector corresponds to the SUGRA potential and when the
hidden sector is correctly stabilized then the quintessence field
acquires a mass of the order of the gravitino mass. If the hidden
sector is more complicated, the runaway shape can be preserved and it
is interesting to see whether local gravity tests can constrain this
type of models. The cosmological implications of the new
potential~(\ref{newpotde}) are worked out in detail in
Sec.~\ref{Implications for Cosmology}.

\subsection{The Observable Sector}

After having described how the dark energy sector looks like for the
SUGRA model, we now study the observable sector. On very general
grounds, it was shown in Ref.~\cite{BMpart} that the interaction
between the dark energy sector and the observable sector implies a
Yukawa like interaction of the form
\begin{equation}
\label{yuk}
m^2\left(\frac{Q}{\mpl}\right)\bar{\Psi}\Psi \, ,
\end{equation}
where $\Psi $ is a fermionic field. Therefore, the above expression
implies that the fermion masses become quintessence field dependent
quantities. As we discuss in the following, this implies a series of
interesting effects as the presence of a fifth force and/or a
violation of the weak equivalence principle. Our goal in this
subsection is to use the general results obtained in
Ref.~\cite{BMpart} and to apply them to the SUGRA model in order to
compute explicitly the functions $m(Q/\mpl)$.

\par

Computing the masses of the fermions first requires to compute the
soft terms. This was done in Ref.~\cite{BMpart}, see
Eqs.~(2.21)--(2.23) of that article, and the general result reads
\begin{eqnarray}
\label{softa} A_{abc} &=&
\lambda _{abc}{\rm e}^{\kappa K_{\rm quint}+\sum _i\vert a_i\vert^2}
\Biggl\{ \biggl(M_{_{\rm S}}+\kappa W_{\rm quint}^{\dagger }\biggr)
+\frac13 \biggl(M_{_{\rm S}}+\kappa W_{\rm quint}^{\dagger
}\biggr)\biggl[\kappa \left(K^{-1}\right)^{d_{\alpha
}^{\dagger}d_{\beta }} \nonumber \\ & & \times \frac{{\partial }K_{\rm
quint}}{\partial d_{\beta }} \frac{{\partial }K_{\rm quint}}{\partial
d_{\alpha }^{\dagger }} +\sum _i\vert a_i\vert ^2-3\biggr] +\frac13
\kappa \left(K^{-1}\right)^{d_{\alpha }^{\dagger}d_{\beta }}
\frac{{\partial }K_{\rm quint}}{\partial d_{\alpha}^{\dagger}}
\frac{{\partial }W_{\rm quint}}{\partial d_{\beta }}\nonumber \\ & &
+\frac13 M_{_{\rm S}}\sum _ia_ic_i\Biggr\}\, ,\\
\label{softb} B_{ab} &=&
\mu _{ab}{\rm e}^{\kappa K_{\rm quint}+\sum _i\vert a_i\vert^2}
\Biggl\{ \biggl(M_{_{\rm S}}+\kappa W_{\rm quint}^{\dagger }\biggr)
+\frac12 \biggl(M_{_{\rm S}}+\kappa W_{\rm quint}^{\dagger
}\biggr)\biggl[\kappa \left(K^{-1}\right)^{d_{\alpha
}^{\dagger}d_{\beta }} \nonumber \\ & & \times \frac{{\partial }K_{\rm
quint}}{\partial d_{\beta }} \frac{{\partial }K_{\rm quint}}{\partial
d_{\alpha }^{\dagger }} +\sum _i\vert a_i\vert ^2-3\biggr] +\frac12
\kappa \left(K^{-1}\right)^{d_{\alpha }^{\dagger}d_{\beta }}
\frac{{\partial }K_{\rm quint}}{\partial d_{\alpha}^{\dagger}}
\frac{{\partial }W_{\rm quint}}{\partial d_{\beta }}\nonumber \\ & &
+\frac12 M_{_{\rm S}}\sum _ia_ic_i\Biggr\}\, ,\\
\label{softm} m_{a\bar{b}} ^2 &=& {\rm e}^{\kappa K_{\rm
quint}+\sum _i\vert a_i\vert^2} \left[M_{_{\rm S}}^2+M_{_{\rm S}}
\left(\kappa W_{\rm quint}+\kappa W_{\rm quint}^{\dagger
}\right)+\kappa ^2W_{\rm quint}W_{\rm quint}^{\dagger }\right]\delta
_{a\bar{b}}\, .
\end{eqnarray}
This is the general form of the soft terms, calculated at the GUT
scale. Then, we have to specialize the above formulas to the dark
sector described by Eqs.~(\ref{quintsector}). Again the fact that
$\mean{W_{\rm quint}}=0$ considerably simplifies the
calculations. Straightforward manipulations lead to
\begin{eqnarray}
\label{softasugra} A_{abc} &=& \lambda _{abc}m_{3/2}^0{\rm e}^{\kappa
K_{\rm quint}}{\rm e}^{\sum _i \vert a_i\vert
  ^2/2}\Biggl[1+\frac13\sum_i \vert a_i\vert^2+\frac13 \sum_i
  a_ic_i+\frac13\biggl(\kappa Q^2+\kappa \xi^2 \nonumber \\ & & +\kappa
  \sum _{\alpha =1}^n\xi _{\alpha }^2-3\biggr)\Biggr]\, ,\nonumber \\ \\
\label{softbsugra} B_{ab} &=& \mu_{ab}m_{3/2}^0{\rm e}^{\kappa K_{\rm
quint}}{\rm e}^{\sum _i\vert a_i\vert^2/2}\Biggl[1+\frac12 \sum
  _i\vert a_i\vert^2+\frac12\sum_i a_ic_i+\frac12\biggl(\kappa
  Q^2+\kappa \xi^2\nonumber \\ & & +\kappa \sum _{\alpha =1}^n\xi
  _{\alpha }^2-3\biggr) \Biggr]\, ,\nonumber \\ \\
\label{softmsugra} m_{a\bar{b}} &=& m_{3/2}^0{\rm
e}^{\kappa K_{\rm quint}/2} \delta _{a\bar{b}}\, ,
\end{eqnarray}
where, again for simplicity, we have assumed $\lambda _{\alpha
}=0$. In the following, one will neglect $\xi $ in comparison with the
vev of $Q$ (as $\xi $ has to be extremely small in order for the
quintessence energy density to be of the order of the critical energy
density today, see above) and one will also consider that $\sum
_{\alpha =1}^n\xi _{\alpha }^2=3/\kappa $. This last relation appears
in the model of the hidden sector discussed before, see
Eq.~(\ref{sumxial}), but is also natural in the present context when
one treats the case where the potential keeps its runaway shape
despite the appearance of the new terms coming from the hidden
sector. This means that $V_{_{\rm DE}}$ should vanish at infinity and,
therefore, that the constant terms cancel in the term $\Upsilon
^4(Q)$. Again, this issue is linked to the cosmological constant
problem.

\par

In order to obtain the masses of the fermions, one should use the
following procedure. In the MSSM, they are two Higgs doublets $H_{\rm
u}$ and $H_{\rm d}$ and the fermions either couple to the Higgs
doublet ``u'' or ``d''. These couplings are different and, therefore,
through the Higgs mechanism, the masses of the fermions either depend
on the vev of the Higgs ``u'' or of the Higgs ``d''. Explicitly, one
has~\cite{BMpart}
\begin{eqnarray}
\label{massfermionu}
m_{\rm u} &=& \lambda _{\rm u}^{_{\rm F}}{\rm e}^{K_{\rm quint}/2+\sum
  _i\vert a_i\vert ^2/2} \frac{v\tan \beta }{\sqrt{1+\tan ^2\beta
  }}\nonumber \\ &=& \lambda _{\rm u}^{_{\rm F}} {\rm e}^{K_{\rm
  quint}/2+\sum _i\vert a_i\vert ^2/2}\left[v(Q)+{\cal
  O}\left(\frac{1}{\tan ^2\beta }\right)\right]\, ,\\ 
\label{massfermiond}
m_{\rm d} &=&\lambda _{\rm d}^{_{\rm F}}{\rm e}^{K_{\rm quint}/2+\sum
  _i\vert a_i\vert ^2/2} \frac{v}{\sqrt{1+\tan ^2\beta }} \nonumber \\
  &=&\lambda _{\rm d}^{_{\rm F}}{\rm e}^{K_{\rm quint}/2+\sum _i\vert
  a_i\vert ^2/2}\left[ \frac{v(Q)}{\tan \beta }+{\cal
  O}\left(\frac{1}{\tan ^2\beta }\right)\right] \, ,
\end{eqnarray}
where $\lambda _{\rm u,d}^{_{\rm F}}$ are the Yukawa coupling which
are, in the minimal setting considered here, independent of the
quintessence field. In the previous equations, $v$ is defined by
$\sqrt{v_{\rm u}^2+v_{\rm d}^2}$, where $v_{\rm u}\equiv v\sin \beta $
and $v_{\rm d}\equiv v\cos \beta $ are the vevs of the Higgs $H_{\rm
u}$ and $H_{\rm d}$ respectively. We have expanded the two previous
expressions in terms of $1/\tan \beta $.  The explicit expression of
$\tan \beta $ reads~\cite{BMpart}
\begin{eqnarray}
\label{tan}
\tan \beta (Q) &=& \frac{2\vert \mu \vert ^2{\rm e}^{\sum _i\vert
    a_i\vert ^2}+m_{H_{\rm u}}^2(Q)+m_{H_{\rm d}}^2(Q)}{2\mu
    B(Q)}\nonumber \\ & & \times \biggl(1 \pm \sqrt{1-4\mu
    ^2B^2(Q)\left[2\vert \mu \vert ^2{\rm e}^{\sum _i\vert a_i\vert
    ^2}+m_{H_{\rm u}}^2(Q)+m_{H_{\rm d}}^2(Q)\right]^{-2}}\biggr)\, .
\end{eqnarray}
Notice that there are two possibilities according to the sign in the
above expression. In the following, we work in the limit of large
$\tan \beta $ and, therefore, take the largest value. In
Eq.~(\ref{tan}), $m^2_{H_{\rm u}}(Q)$ and $m^2_{H_{\rm d}}(Q)$ are the
two loops renormalized Higgs masses given by~\cite{Brax,BMpart}
\begin{eqnarray}
\label{mhu}
m^2_{H_{\rm u}}\left( Q\right) &=& m^2_{H_{\rm d}}(Q)
-0.36\left(1+\frac{1}{\tan^2 \beta}\right)
\Biggl\{\left(m_{3/2}^0\right)^2\left(1-
\frac{1}{2\pi}\right)+8\left(m_{1/2}^0\right)^2 \nonumber \\ & & +
\left(0.28 -\frac{0.72}{\tan ^2\beta }\right )\left[A(Q) + 2
  m^0_{1/2}\right]^2\Biggr\}\, ,\\
\label{mhd}
m^2_{H_{\rm d}}\left(Q\right) &=&
      \left(m_{3/2}^0\right)^2\left(1-\frac{0.15}{4\pi}\right) +
      \frac{1}{2} \left(m^0_{1/2}\right)^2\, .
\end{eqnarray}
We see from the two above expressions that the quintessence field
dependence of the fermions masses is controlled by two functions,
$A(Q)$ and $B(Q)$. The explicit expressions of these soft terms, see
Eqs.~(\ref{softasugra})--(\ref{softmsugra}), allow us to extract the
functions $A(Q)$ and $B(Q)$ by $A_{abc}\equiv {\rm e}^{\kappa K_{\rm
quint}} A(Q)\lambda _{abc}$ and $B_{ab}\equiv {\rm e}^{\kappa K_{\rm
quint}} \mu B(Q)\epsilon _{ab}$. If we restrict our considerations to
$a_i=0$, which is a case of interest since this is at the same time
compatible with $\mean{z_i}\ll m_{_{\rm C}}$ and with a possible
runaway potential (using a non trivial dependence of the coefficient
$c_i$, see the discussions before), one arrives at
\begin{equation}
A(Q)=M_{_{\rm S}}\left(1+\frac{\kappa Q^2}{3}\right)\, ,\quad
B(Q)=M_{_{\rm S}}\left(1+\frac{\kappa Q^2}{2}\right)\, .
\end{equation}
Notice that $A$ and $B$ follow the same universal relationship as in
the mSUGRA model despite the presence of the quintessence field. Then,
one deduces that $\tan \beta (Q)$ in the SUGRA model can be expressed
as
\begin{equation}
\label{tanbetasugra}
\tan \beta(Q) \simeq \frac{\delta _1+\delta _2\kappa Q^2+ \delta_3
  \kappa^2 Q^4}{\delta
_4+\delta_5 \kappa Q^2}\left[1+\sqrt{1+\frac{\left(\delta _4+\delta
_5\kappa Q^2\right)^2}{\left(\delta _1+\delta _2\kappa
Q^2+ \delta_3 \kappa^2 Q^4\right)^2}}\right]\, ,
\end{equation}
where the coefficients $\delta _1$, $\delta _2$, $\delta _3$,
$\delta_4$ and $\delta _5$ can easily be evaluated in terms of the
physical parameters characterizing the model from the previous
equations, \ie $\mu$, $m_{3/2}^0$ and $m_{1/2}^0$ given at the GUT
scale. As already mentioned, this result is valid both in the case
where the hidden sector fields are stabilized with $a_i=c_i=0$ and
when the hidden sector dynamics is tuned to reach $a_i=0$ and $c_i\ne
0$ possibly leading to a runaway potential. In the latter the role of
the gravitational tests is crucial in discriminating models.

\par

The expression of the scale $v(Q)$ can also be obtained from the
minimization of the Higgs potential along the lines described in
Ref.~\cite{BMpart}. One obtains
\begin{equation}
\label{v}
v(Q)=\frac{2}{\sqrt{g^2+g'{}^2}}{\rm e}^{\kappa K_{\rm quint}/2}
\sqrt{\left\vert \left \vert \mu \right \vert ^2+m_{H_{\rm
u}}^2\right\vert} +{\cal O}\left(\frac{1}{\tan \beta }\right)\, ,
\end{equation}
where as before, we have used $a_i=0$. The value of $v$ today is known
and is $v\sim 174 \, \mbox{GeV}$. Therefore, recalling that
$m_{Z^0}^2=(g^2+g'{}^2)v^2/2$ with $m_{Z^0}\sim 91.6 \, \mbox{GeV}$
and expressing Eq.~(\ref{v}) at vanishing redshift allows us to
determine the $\mu $ parameter. Explicitly, one has
\begin{equation}
\label{mu}
\left\vert \mu \right \vert =\sqrt{\frac12 m_{Z^0}^2 {\rm e}^{-\kappa
\left(Q_{\rm now}^2+\xi ^2+\sum _{\alpha }\xi _{\alpha
}^2\right)}-m_{H_{\rm u}}^2}\, .
\end{equation}
In the following, as already explained, we neglect $\xi $ and use
$\sum _{\alpha =1}^n\xi _{\alpha }^2=3/\kappa $. The above formula
also depends on the value of the vev of the quintessence field today,
$Q_{\rm now}$. This one should be determined by the cosmological
evolution of the field in the potential $V_{_{\rm DE}}$ and is
therefore fixed once the parameters (\ie $m_{3/2}^0$, $m_{1/2}^0$ etc
\dots ) have been chosen. However, in the runaway case, it is clear
that one always have $\kappa^{1/2} Q_{\rm now}\sim 1$ and, for
definiteness, we will take $\kappa ^{1/2}Q_{\rm now}=1$ in the
following, the final results being (almost) independent of the precise
value of $\kappa^{1/2} Q_{\rm now}$.

\par

\begin{figure}
\begin{center}
\includegraphics[width=14cm]{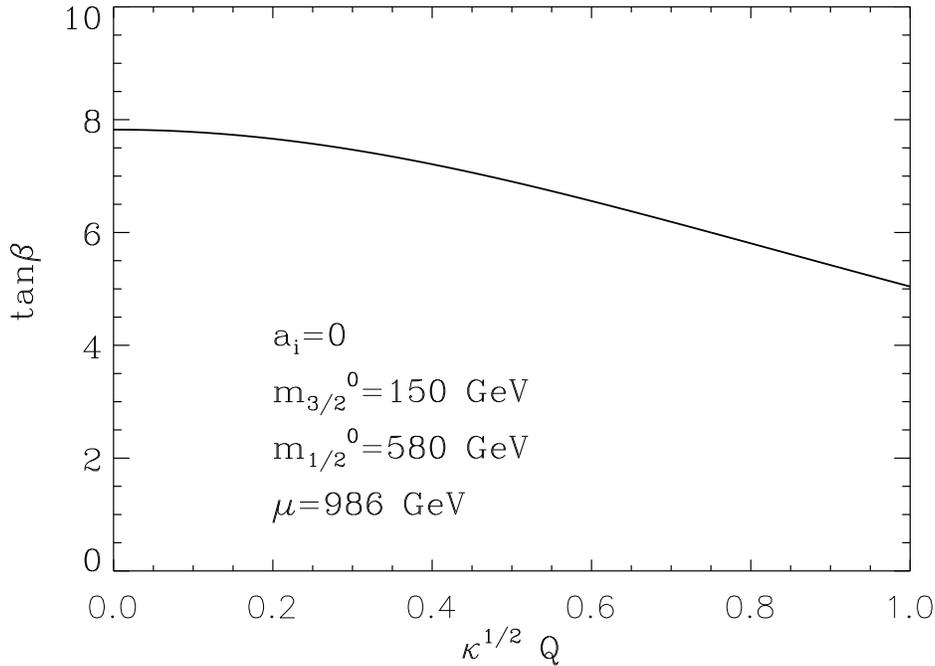}
\caption[]{Evolution of $\tan \beta $ versus the vev of the
quintessence field according to Eq.~(\ref{tan}) or
Eq.~(\ref{tanbetasugra}). }
\label{tanbeta}
\end{center}
\end{figure}

\begin{figure}
\begin{center}
\includegraphics[width=14cm]{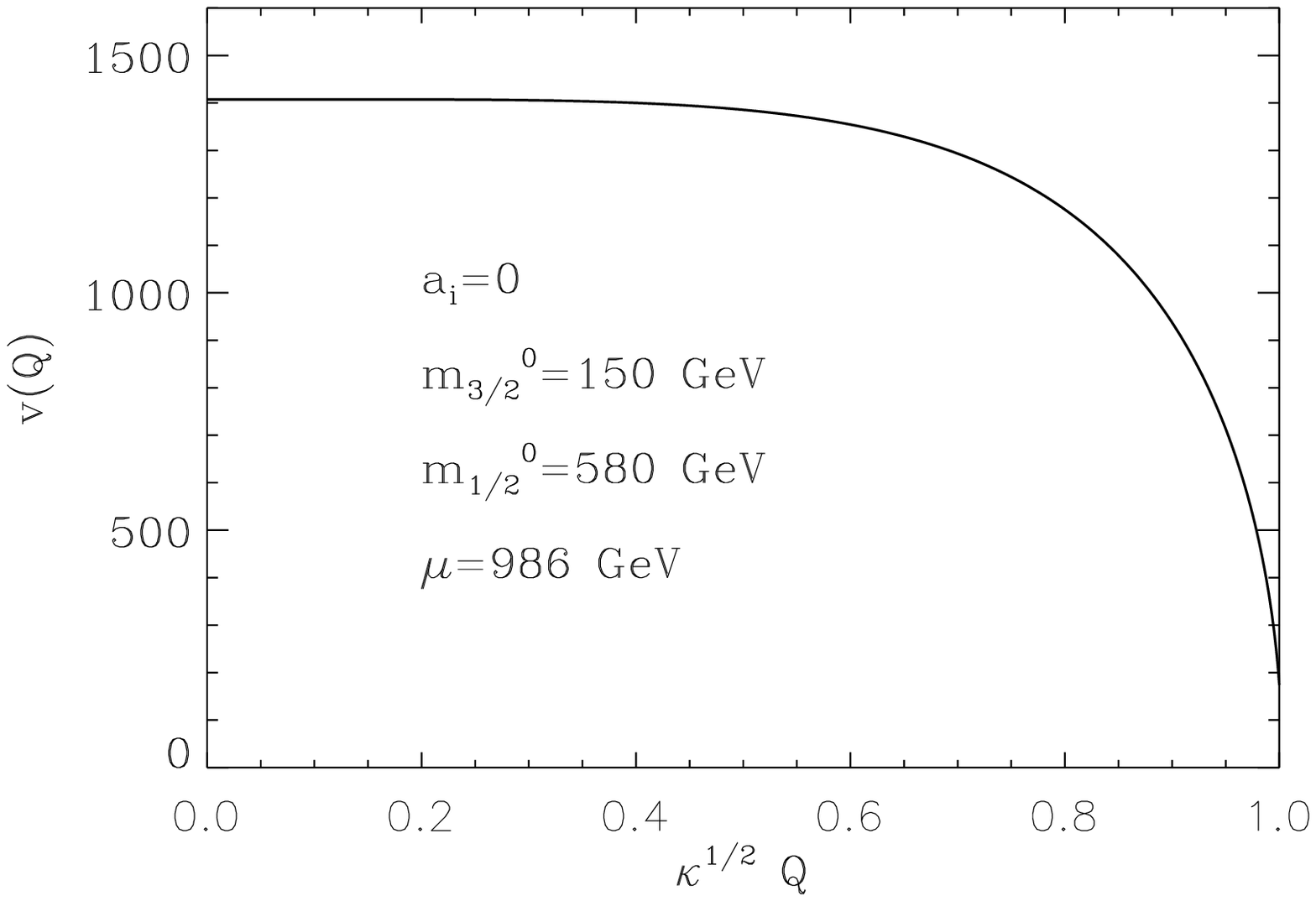}
\caption[]{Evolution of  $v$ versus the vev of
  the quintessence field as predicted by Eq.~(\ref{v})}.
\label{vplot}
\end{center}
\end{figure}

Let us also remark that the above formula giving $\tan \beta $ and $v$
are only approximated formulas and, as announced above, valid only
when terms like $1/\tan ^2 \beta $ are negligible in Eqs.~(\ref{mhu})
and~(\ref{mhd}). Otherwise one would have to deal with a
transcendental equation. If necessary, this equation can always be
solved numerically, but, in this article, we always consider the
approximation where the various quantities of interest are expanded in
$1/\tan \beta $. The corresponding evolution of $\tan \beta $ as given
by Eq.~(\ref{tanbetasugra}) is represented in Fig.~\ref{tanbeta}.
We see that $\tan \beta $ does not change sign, which would have been
problematic once the sign of $\mu$ is fixed. In Fig.~\ref{tanbeta},
one check that the electroweak symmetry breaking conditions are indeed
satisfied in the whole range for which $\tan \beta $ is plotted. We
mentioned before that this plot has been obtained by neglecting the
terms $1/\tan ^2\beta $ in Eqs.~(\ref{mhu}) and (\ref{mhd}). One can
verify that, for our choice of parameters, this is a good
approximation since $\tan \beta \gtrsim 5$. In Fig.~\ref{vplot}, we
have also represented the scale $v(Q)$ given by Eq.~(\ref{v}) for the
same values of the parameters as before.

\par

\begin{figure}
\begin{center}
\includegraphics[width=14cm]{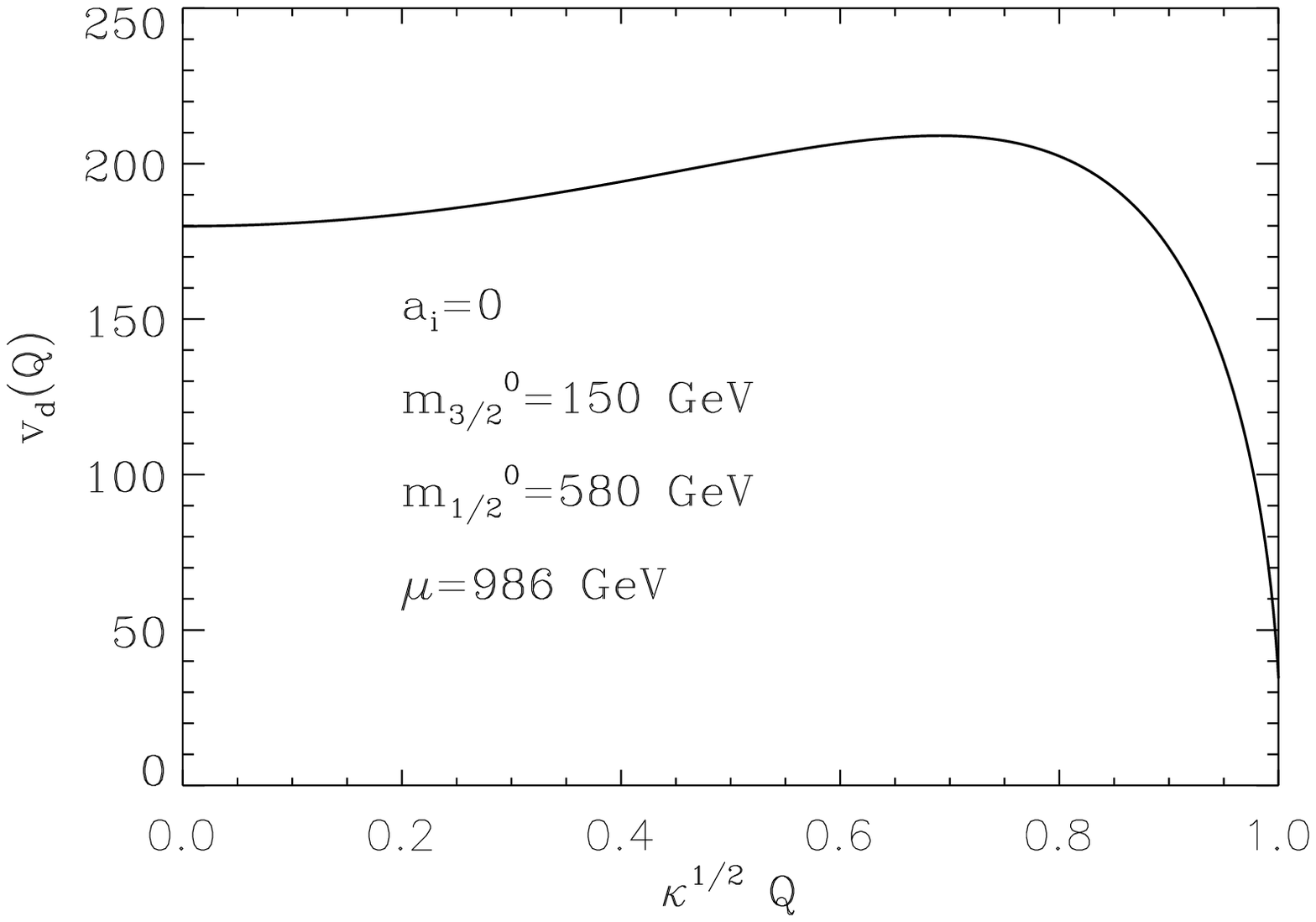}
\caption[]{Evolution of the Higgs vev $v_{\rm d}$ versus the vev of
  the quintessence field as predicted by the second of
  Eqs.~(\ref{massfermiond}).}
\label{vdplot}
\end{center}
\end{figure}

If the expressions of $\tan \beta $ and $v(Q)$ are used in
Eqs.~(\ref{massfermionu}) and~(\ref{massfermiond}), this gives $m_{\rm
u, \rm d}(Q)$. To our knowledge, this is the first time that the $Q$
dependence of the fermions masses is calculated in a precise model
from first principles. The vev $v_{\rm u}(Q)$ is equal to $v(Q)$ at
leading order and the vev $v_{\rm d}(Q)$ is represented in
Fig.~\ref{vdplot}.

\par

Finally, let us remark that, very often in the literature, the
function $m_{\rm u,d}(Q)$ is just postulated see, for instance,
Ref.~\cite{Khoury} where $m_{\rm u,d}(Q)\propto \exp\left(\beta \kappa
^{1/2}Q/2\right)$, $\beta $ being a free parameter. We see that, even
in our oversimplified model, the dependence can be much more
complicated.

\section{Implications for Gravity Experiments}
\label{Implications for Gravity Experiments}

\subsection{Fifth Force Constraints}
\label{Fifth Force Constraints}

In this section, we study the consequences of the fact that the
fermions masses are now $Q$-dependent quantities. It is known that
this can cause serious problems coming from gravitational experiments
since this implies the presence of a fifth force and a violation of
the weak equivalence principle, see the next subsection. In fact, this
crucially depends on the mass of the quintessence field. If the mass
of the quintessence field is larger than $10^{-3} \mbox{eV}$, then the
gravitational constraints are always satisfied as the range of the
force mediated by $Q$ is less than one millimeter. We see that this
occurs when the functions $a_i(Q)$ and $c_i(Q)$ vanish or are
constant. Indeed, we have shown that, in this case, the potential has
a minimum and acquires a mass $\sim m_{3/2}^0\gg
10^{-3}\mbox{eV}$. Therefore, we reach the conclusion that the SUGRA
model, with a hidden sector such that the fields are correctly
stabilized, is free from gravitational problems. As discussed in the
next section, the problems rather originate from cosmological
considerations.

\par

On the other hand, if the mass of the quintessence field is less than
$10^{-3} \mbox{eV}$, the range of the quintessence field is large and
generically there will be violations of the weak equivalence principle
and a large fifth force. In the case of the SUGRA model, this requires
non trivial functions $a_i(Q)$ and $c_i(Q)$. In this situation, in
order to avoid fifth force experiments such as the recent Cassini
spacecraft experiment, one must require that the Eddington
(post-Newtonian) parameter $\vert \gamma -1\vert \le 5\times 10^{-5}$,
see Ref.~\cite{GR}. The Eddington parameter $\gamma $ is related to
the parameters $\alpha _{\rm u,d}$ by $\gamma =1+\alpha _{\rm u,d}^2$,
where $\alpha _{\rm u,d}$ can be expressed as
\begin{equation}
\label{alpha}
\alpha_{\rm u,d}(Q) \equiv \left\vert \frac{1}{\kappa^{1/2}}
\frac{{\rm d}\ln m_{\rm u,d}^{_{\rm F}}(Q)}{{\rm d} Q}\right \vert
=\left \vert \frac{1}{\kappa^{1/2}} \frac{{\rm d}\ln \left[{\rm
e}^{K_{\rm quint}/2+\sum _i\vert a_i\vert ^2/2} v_{\rm
u,d}(Q)\right]}{{\rm d} Q}\right \vert \, ,
\end{equation}
Therefore, the coefficients $\alpha _{\rm u,d}$ can be determined
explicitly from the formulas giving $m_{\rm u,d}$. The model is free
from difficulties if $\alpha_{\rm u,d}^2\le 10^{-5}$.  Clearly, the
coefficients can be calculated explicitly as soon as the functions
$a_i(Q)$ and $c_i(Q)$ are known. Here, we perform this calculation for
the choice $a_i=0$ leaving the function $c_i\ne 0$ free to lead to a
runaway potential as already discussed at length before. In addition,
in order to deal with the simplest model, we consider that the scale
$M_{_{\rm S}}$ does not dependent on $Q$ as indicated by the model of
SUSY breaking presented before. One obtains
\begin{eqnarray}
\label{alphadef} 
\alpha _{\rm u} &=&\frac{\kappa ^{1/2}}{2}\partial _Q K_{\rm
quint}+\frac{\kappa ^{-1/2}}{\tan \beta \left(1+\tan ^2\beta \right)}
\frac{{\rm d}\tan \beta }{{\rm d}Q}+\frac{\kappa ^{-1/2}}{v}\frac{{\rm
d}v}{{\rm d}Q} \nonumber \\ &=& \frac{\kappa ^{1/2}}{2}\partial _Q
K_{\rm quint} +\frac{\kappa ^{-1/2}}{v}\frac{{\rm d}v}{{\rm d}Q}
+{\cal O}\left(\frac{1}{\tan ^2\beta }\right) \, ,\\ \alpha _{\rm d}
&=&\frac{\kappa ^{1/2}}{2}\partial _Q K_{\rm quint}- \frac{\kappa
^{-1/2}\tan \beta }{1+\tan ^2\beta } \frac{{\rm d}\tan \beta }{{\rm
d}Q}+\frac{\kappa ^{-1/2}}{v}\frac{{\rm d}v}{{\rm d}Q} \nonumber \\
&=& \frac{\kappa ^{1/2}}{2}\partial _Q K_{\rm quint}-\frac{\kappa
^{-1/2}}{\tan \beta }\frac{{\rm d}\tan \beta }{{\rm d}Q} +\frac{\kappa
^{-1/2}}{v}\frac{{\rm d}v}{{\rm d}Q} +{\cal O}\left(\frac{1}{\tan
^2\beta }\right)\, ,
\end{eqnarray}
where the derivative of the function $\tan \beta (Q)$ can be expressed
as
\begin{eqnarray}
\frac{{\rm d}\tan \beta }{{\rm d}Q} &=& \left(\frac{{\rm d}m_{\rm
    H_{\rm u}}^2}{{\rm d}Q}+\frac{{\rm d}m_{\rm H_{\rm d}}^2}{{\rm
    d}Q}\right)\left(2\vert\mu \vert ^2 +m_{\rm H_{\rm u}}^2+m_{\rm
    H_{\rm d}}^2\right)^{-1}\tan \beta -\frac{1}{B(Q)}\frac{{\rm
    d}B(Q)}{{\rm d}Q}\tan \beta \nonumber \\ & & \pm 2\mu
    \left(2\vert\mu \vert ^2 +m_{\rm H_{\rm u}}^2+m_{\rm H_{\rm
    d}}^2\right)^{-1} \Biggl[1-4\mu ^2B^2(Q)\biggl(2\vert\mu \vert ^2
    +m_{\rm H_{\rm u}}^2 +m_{\rm H_{\rm
    d}}^2\biggr)^{-2}\Biggr]^{-1/2}\nonumber \\ & & \times
    \Biggl[-\frac{{\rm d}B(Q)}{{\rm d}Q} +B(Q)\left(\frac{{\rm
    d}m_{\rm H_{\rm u}}^2}{{\rm d}Q}+\frac{{\rm d}m_{\rm H_{\rm
    d}}^2}{{\rm d}Q}\right)\biggl(2\vert\mu \vert ^2 +m_{\rm H_{\rm
    u}}^2+m_{\rm H_{\rm d}}^2\biggr)^{-1}\Biggr]\, , \\ &\sim &
    \left(\frac{{\rm d}m_{\rm H_{\rm u}}^2}{{\rm d}Q}+\frac{{\rm
    d}m_{\rm H_{\rm d}}^2}{{\rm d}Q}\right)\left(2\vert\mu \vert ^2
    +m_{\rm H_{\rm u}}^2+m_{\rm H_{\rm d}}^2\right)^{-1}\tan \beta
    -\frac{{\rm d}\ln B(Q)}{{\rm d}Q}\tan \beta \, .
\end{eqnarray}
The last expression is valid at leading order and we use the explicit
form of the functions $A(Q)$ and $B(Q)$ for the SUGRA model in order
to evaluate the derivatives of the soft terms 
\begin{eqnarray}
\frac{{\rm d}A(Q)}{{\rm d}Q}=\frac{2m_{3/2}^0}{3}\kappa Q\, , \quad
    \frac{{\rm d}B(Q)}{{\rm d}Q}=m_{3/2}^0\kappa Q\, ,
\end{eqnarray}
since $M_{_{\rm S}}$ is constant. Consequently, the derivatives of the
Higgs masses can be expressed as
\begin{eqnarray}
\frac{{\rm d}m_{\rm H_{\rm u}}^2}{{\rm d}Q}\simeq -0.72\times 0.28
\frac{{\rm d}A(Q)}{{\rm d}Q}\left[A(Q)+2m^0_{1/2}\right]\, , \quad
\frac{{\rm d}m_{\rm H_{\rm d}}^2}{{\rm d}Q}\simeq 0\, ,
\end{eqnarray}
the symbol ``approximate'' in the last two equations meaning that we
have used the fact that the terms in $1/\tan ^2\beta $ have been
neglected in the expression of the above formulas. In the previous
calculation, we have also used the fact that $m_{1/2}^0$ is
constant. This means that we have assumed specific forms for the gauge
functions $f_{_{G}}$, namely we have considered that they do not
depend on $Q$ and $z_i$ but only on the dark sector fields $X_{\alpha
}$. Then, if we parameterize $\xi _{\alpha }$ and the derivative of
$f=f_{_{G}}$ as
\begin{equation}
\xi_\alpha= \sqrt \frac{ 3}{ \kappa} e_\alpha\, ,\quad \kappa
 ^{-1/2}\frac{\partial f}{\partial X_\alpha} = h_\alpha \, ,
\end{equation}
where the coefficients $e_{\alpha }$ and $h_{\alpha }$ are of order
one, one finds that
\begin{equation}
m_{1/2}^0= \sqrt{3}m_{3/2}^0\sum_\alpha e_\alpha h_\alpha\, ,
\end{equation}
with no dependence on $Q$ and a model dependent prefactor of
$m_{3/2}^0$.

\par

\begin{figure*}
\includegraphics[width=14cm]{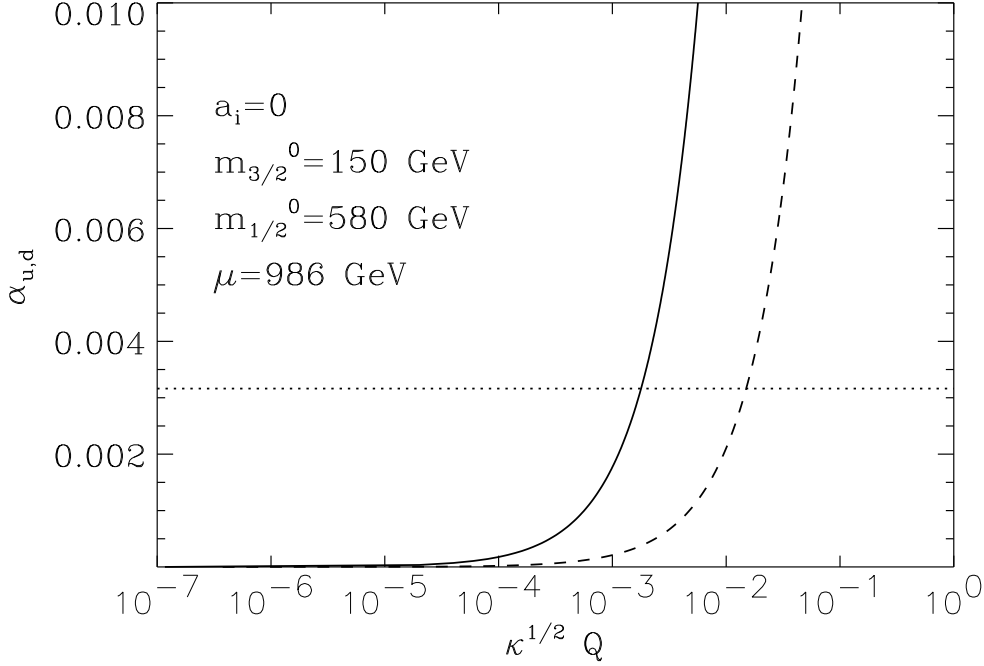}
\caption[]{Evolution of the coefficients $\alpha _{\rm u}$ (solid
line) and $\alpha _{\rm d}$ (dashed line) versus the vev of the
quintessence field.}
\label{alphaplot}
\end{figure*}

As an example we have plotted in Fig.~\ref{alphaplot} the
gravitational coupling constants $\alpha_{\rm u, \rm d}$ for a
realistic situation where $a_i=0$ and the parameters $m_{3/2}^0=150\,
\mbox{GeV}$, $m_{1/2}^0=580\, \mbox{GeV}$ and, hence, $\mu =986\,
\mbox{GeV}$, that is to say $m_{3/2}^0$ and $m_{1/2}^0$ roughly
speaking of the same order of magnitude as indicated by the previous
calculation.  We see that the limit $\alpha _{\rm d}\sim 10^{-2.5}$ is
reached for relatively large value of the quintessence field vev, or
the order of $\kappa^{1/2} \mean{Q}\sim 10^{-4}-10^{-2}$. This implies
that the SUGRA model with $a_i=0$ and $c_i\ne 0$ to obtain a runaway
potential with $Q_{\rm now}\sim \mpl$ is excluded unless the
dependence of the masses on $Q$ in (\ref{alpha}) involves a $Q$
dependent Yukawa coupling compensating exactly the $Q$ dependence of
the Higgs vevs.  Although this is not excluded, this is a functional
fine--tuning which is hard to explain.

\subsection{Violation of the Weak Equivalence Principle}
\label{Violation of the Equivalence Principle}

As explained in detail in Ref.~\cite{BMpart}, the fact that, in the
MSSM, the fermions couple differently to the two Higgs doublets
$H_{\rm u}$ and $H_{\rm d}$ causes, in the presence of dark energy, a
violation of the weak equivalence principle. This violation is
quantified in terms of the $\eta_{_{\rm AB}}$ parameter defined
by~\cite{DP,damour,JP}
\begin{equation}
\eta_{_{\rm AB}} \equiv \left(\frac{\Delta a}{a}\right)_{_{\rm AB}}
=2\frac{a_{_{\rm A}}-a_{_{\rm B}}}{a_{_{\rm A}}+a_{_{\rm B}}}\, ,
\end{equation}
for two test bodies A and B in the gravitational background of a third
one E. Current limits~\cite{Su} give $\eta_{_{\rm AB}} =(+0.1\pm
2.7\pm 1.7)\times 10^{-13}$. The $\eta _{_{\rm AB}}$ parameter was
computed in Ref.~\cite{BMpart} where a general formula was
derived. Applying this general result to the case of the SUGRA
potential leads to
\begin{eqnarray}
\eta_{_{\rm AB}} &=& \frac{1}{2}\kappa ^{-1/2}\alpha _{_{\rm
E}}\biggl[ \frac{\partial }{\partial Q}\left(\frac{\sigma '}{\Lambda
_{_{\rm QCD}}}\right)\left(\frac{N_{_{\rm A}}+Z_{_{\rm A}}}{M_{_{\rm
A}}}-\frac{N_{_{\rm B}}+Z_{_{\rm B}}}{M_{_{\rm B}}}\right) \nonumber
\\ & & +\frac{\partial }{\partial Q}\left(\frac{\delta '}{\Lambda
_{_{\rm QCD}}}\right)\left(\frac{N_{_{\rm A}}-Z_{_{\rm A}}}{M_{_{\rm
A}}} -\frac{N_{_{\rm B}}-Z_{_{\rm B}}}{M_{_{\rm B}}}\right) \biggr]\,
,
\end{eqnarray}
where $N_{_{\rm A}}$ (respectively $N_{_{\rm B}}$) represents the
number of neutrons in the atom A, the body A, by definition, being
made of this type of atoms (respectively in the atom B) and $Z_{_{\rm
A}}$ (respectively $Z_{_{\rm B}}$) represents the number of protons in
the atom A (respectively in the atom B). The coefficients $\alpha
_{_{\rm E}}$ is defined by $\alpha _{_{\rm E}}\equiv \kappa
^{-1/2}{\rm d}\ln m_{_{\rm E}}/{\rm d}Q$ but for two pairs of test
bodies, the ratio $\eta_{_{\rm AB}}/\eta_{_{\rm BC}}$ is independent
of the background object E. The quantity $\Lambda _{_{\rm QCD}}\sim
180 \mbox{MeV}$ is the QCD scale. Finally the variation of the
coefficients $\sigma '$ and $\delta '$ can be expressed as
\begin{eqnarray}
\frac{\partial }{\partial Q}\left(\frac{\sigma '}{\Lambda _{_{\rm
QCD}}}\right) &=& \frac12 \frac{\kappa ^{1/2}}{\Lambda _{_{\rm QCD}}}
\left(b_{\rm u}+b_{\rm d}\right)\left(\alpha _{\rm u}m_{\rm u}+\alpha
_{\rm d}m_{\rm d}\right) +\frac12 \frac{\kappa ^{1/2}}{\Lambda _{_{\rm
QCD}}}\alpha _{\rm d}m_{\rm e}\, , \\ \frac{\partial }{\partial
Q}\left(\frac{\delta '}{\Lambda _{_{\rm QCD}}}\right) &=& -\frac12
\frac{\kappa ^{1/2}}{\Lambda _{_{\rm QCD}}} \left(b_{\rm u}-b_{\rm
d}\right)\left(\alpha _{\rm u}m_{\rm u}-\alpha _{\rm d}m_{\rm
d}\right) -\frac12 \frac{\kappa ^{1/2}}{\Lambda _{_{\rm QCD}}}\alpha
_{\rm d}m_{\rm e}\, ,
\end{eqnarray}
where $b_{\rm u} + b_{\rm d}\sim 6$, $b_{\rm u}-b_{\rm d}\sim 0.5$,
$\sigma'/\Lambda_{_{\rm QCD}}\sim 3.8\times 10^{-2}$,
$\delta'/\Lambda_{_{\rm QCD}}\sim 4.2 \times 10^{-4}$, $m_{\rm u}\sim
5 \mbox{MeV}$, $m_{\rm d}\sim 10 \mbox{MeV}$ and $m_{\rm e}\sim 0.5
\mbox{MeV}$. Let us notice that the coefficient $\alpha _{\rm d}$
appears in front of the mass of the electron $m_{\rm e}$ because, in
the MSSM, the electron behaves as a ``d'' particle. If we compare the
above equations with the ones of Ref.~\cite{BMpart}, one sees that
there is no variations of the fine structure constant and/or of the
gauge function $f_{_{\rm QCD}}$. This is simply because, in this
article, we assume that they are constant.

\par

Let us recall that when the mass of the quintessence field is of the
order of the gravitino mass, all the gravity tests, among which is the
equivalence principle violation are trivially satisfied. However, it
is interesting to compute the $\eta _{_{\rm AB}}$ in the SUGRA model
with the non trivial hidden sector. One obtains
\begin{eqnarray}
\eta _{_{\rm AB}} &\sim & \frac{1}{2}\alpha _{_{\rm E}}\left(0.084
  \alpha _{\rm u}+0.168 \alpha _{\rm d}\right) \left(\frac{N_{_{\rm
  A}}+Z_{_{\rm A}}}{M_{_{\rm A}}}-\frac{N_{_{\rm B}}+Z_{_{\rm
  B}}}{M_{_{\rm B}}}\right) + \frac{1}{2}\alpha _{_{\rm
  E}}\left(-0.0069 \alpha _{\rm u}+0.013 \alpha _{\rm d}\right)
  \nonumber \\ & & \times \left(\frac{N_{_{\rm A}}-Z_{_{\rm
  A}}}{M_{_{\rm A}}}-\frac{N_{_{\rm B}}-Z_{_{\rm B}}}{M_{_{\rm
  B}}}\right)\, .
\end{eqnarray}
when $a_i=0$ and $c_i\ne 0$ to lead to a runaway potential. It is
interesting to notice that the calculations of the equivalence
principle violation is directly related to the supergravity Lagrangian
and, therefore, originates from first principle. A good order of
magnitude estimate of the parameter $\eta _{_{\rm AB}}$ is simply, see
Fig.~\ref{alphaplot}
\begin{equation}
\eta _{_{\rm AB}}\sim \alpha _{\rm u}^2\, ,
\end{equation}
where we have assumed $\alpha _{_{\rm E}}\sim \alpha _{\rm u}\gg
\alpha _{\rm d}$. In order to comply with the currently available
experimental limits this implies
\begin{equation}
\kappa ^{1/2}Q_{\rm now}\ll 10^{-6}\, .
\end{equation}
The above number is obtained for $m_{3/2}^0=150\, \mbox{GeV}$ and
$m_{1/2}^0=580\, \mbox{GeV}$, which implies $\mu =985\, \mbox{GeV}$,
\ie the values used in Fig.~\ref{alphaplot}. This means that
constraints on the weak equivalence principle are able to rule out
this version of the SUGRA model. In other words, if the hidden sector
is chosen such that it leads to an interesting cosmological model,
then local tests of gravity are able to kill the scenario (at least
for these values of the parameters).

\subsection{The Proton-Electron Mass Ratio}
\label{The Proton-Electron Mass Ratio}

Another consequence of coupling the MSSM to dark energy is that this
will cause the proton to electron mass ratio, $r\equiv m_{\rm
p}/m_{\rm e}$, to vary with time, that is to say with the
redshift. This is an important consequence since, experimentally,
there is an indication for a possible variation, $\Delta r/r\sim
(2.0\pm 0.6) \times 10^{-5}$~\cite{reinhold}. The proton mass can be
written as
\begin{equation}
m_{\rm p}= C_{_{\rm QCD}}\Lambda _{_{\rm QCD}}+ b_{\rm u} m_{\rm u} +
b_{\rm d} m_{\rm d} + C_{\rm p} \alpha_{_{\rm QED}}\, ,
\end{equation}
where $m_{\rm u}$ and $m_{\rm d}$ are the mass of the u and d quarks,
$C_{_{\rm QCD}}\sim 5.2$ and $C_{\rm p}\alpha _{_{\rm QED}}\sim
0.63\mbox{MeV}$. Recalling that the electron behaves as a ``d''
particle, this leads to
\begin{equation}
r=b_{\rm u}\frac{\lambda _{\rm u}}{\lambda _{\rm e}}\tan \beta +b_{\rm
d}\frac{\lambda _{\rm d}}{\lambda _{\rm e}} +\sqrt{1+\tan ^2\beta
}\frac{C_{_{\rm QCD}}\Lambda _{_{\rm QCD}}+C_{\rm p}\alpha _{_{\rm
QED}}}{\lambda _{\rm e}v(Q)}{\rm e}^{-\kappa K_{\rm quint}/2}\, ,
\end{equation}
where $\lambda _{\rm u}$, $\lambda _{\rm d}$ and $\lambda _{\rm e}$
are respectively the Yukawa couplings of the quarks u, d and of the
electron. We also assume, for simplicity, that the gauge functions are
constant, see above. Using Eqs.~(\ref{massfermionu})
and~(\ref{massfermiond}), one can express the Yukawa couplings in
terms of the fermion masses at present time, generically denoted
``$m^0$''. This leads to
\begin{eqnarray}
r &=& b_{\rm u}\frac{m_{\rm u}^0}{m_{\rm e}^0}\frac{\tan \beta
(Q)}{\tan \beta (Q_{\rm now})}+b_{\rm d}\frac{m_{\rm d}^0}{m_{\rm
e}^0} + \frac{C_{_{\rm QCD}}\Lambda _{_{\rm QCD}}+C_{\rm p}\alpha
_{_{\rm QED}}}{m _{\rm e}^0} \sqrt{\frac{\left \vert \mu \right \vert
^2+m_{H_{\rm u}}^2(Q_{\rm now})}{\left \vert \mu \right \vert
^2+m_{H_{\rm u}}^2(Q)}}\nonumber \\ & & \times \frac{\tan \beta
(Q)}{\tan \beta (Q_{\rm now})} \, .
\end{eqnarray}
As before, in order to be consistent, we have to expand the above
equation in terms of $1/\tan \beta $. Then, one obtains
\begin{equation}
r\sim r_{\rm now}+\left (b_{\rm u}\frac{m_{\rm u}^0}{m_{\rm e}^0}
+\frac{C_{_{\rm QCD}}\Lambda _{_{\rm QCD}}+C_{\rm p}\alpha _{_{\rm
QED}}}{m _{\rm e}^0} \right ) \left[\frac{\tan \beta (Q)}{\tan \beta
(Q_{\rm now})}-1\right]\, .
\end{equation}
One can also Taylor expand the tangent function and use the expression
of the derivative of the tangent at leading order.  This gives
\begin{equation}
\frac{\Delta r}{r}\sim \left[\frac{1}{\kappa ^{1/2}}\left(\frac{{\rm
d}m_{\rm H_{\rm u}}^2}{{\rm d}Q}+\frac{{\rm d}m_{\rm H_{\rm
d}}^2}{{\rm d}Q}\right)\biggl(2\vert\mu \vert ^2 +m_{\rm H_{\rm
u}}^2+m_{\rm H_{\rm d}}^2\biggr)^{-1}-\frac{1}{\kappa
^{1/2}}\frac{{\rm d}\ln B}{{\rm d}Q} \right ]\kappa ^{1/2}\Delta Q \,
.
\end{equation}
For instance, if one uses our fiducial model with $m_{3/2}^0=150
\mbox{GeV}$ and $m_{1/2}^0=580 \mbox{GeV}$, one has $\mu =985.7 \,
\mbox{GeV}$, $m_{\rm H_{\rm u}}^2\sim -971650.6\, \mbox{GeV}^2$ and
$m_{\rm H_{\rm d}}^2\sim 190431.3\, \mbox{GeV}^2$ where we recall, for
order of magnitude estimate, we have taken $\kappa ^{1/2}Q_{\rm
now}=1$ which is consistent with a runaway behavior. This leads to
\begin{equation}
\left\vert \frac{\Delta r}{r}\right\vert \sim 0.69\kappa ^{1/2}\Delta
Q\, ,
\end{equation}
This means that, in order to comply with the recent bound, one should
have
\begin{equation}
\kappa ^{1/2}\Delta Q<10^{-5}\, ,
\end{equation}
in the range $z\in [0,3]$. One sees that this constraint is less
strong than the one obtained from the equivalence principle, at least
for this choice of parameters, but is also sufficient to rule out the
model.

\subsection{Cold Dark and Baryonic Energy Densities}
\label{Cold Dark and Baryonic Energy Densities}

The coupling between the observable sector (radiation and matter) and
the quintessence sector also affects the way the energy densities
scale with the Friedmann-Lemaitre-Robertson-Walker (FLRW) scale factor
$a$. In a minimal setting, it turns out the quintessence field couples
to matter and not to radiation since the gauge functions $f_{_{G}}$
are chosen to be constant. This implies that the radiation energy
density still behaves like
\begin{equation}
\rho_{\rm rad} \propto \frac{1}{a^4}\, .
\end{equation}
In particular, in this case, there is no variation of the
fine--structure constant. On the contrary since the masses of the
particles depend on $\kappa Q^2$, the matter density behaves as
\begin{equation}
\label{rhomat}
\rho_{\rm mat} =\sum _a n_a m_{\rm u,d}^{_{\rm F}}\left(\kappa
Q^2\right)\, ,
\end{equation}
where $m_{\rm u,d}^{_{\rm F}}(\kappa Q^2)$ is the mass of non
relativistic species the expression of which has been obtained
before. The quantity $n_a$ is the number of non-relativistic particles
which is conserved $ \dot n_a +3H n_a=0 $, where $H=\dot{a}/a$ is the
Hubble parameter. This leads to
\begin{equation}
\label{rhomat2}
\rho_{\rm mat}\propto \frac{1}{a^3}\sum _a n_a^0 m_{\rm u,d}^{_{\rm
F}}\left(\kappa Q^2\right)\, ,
\end{equation}
where $n_a^0$ is just a constant. Again, in the most general case, the
scaling of the cold dark and baryonic energy densities depends on the
functions $a_i(Q)$ and $c_i(Q)$ and is completely explicit in the case
$a_i=0$. Let us also notice that cosmological consequences of
Eq.~(\ref{rhomat2}) have been studied in Ref.~\cite{Khoury}. As one
can see, the coupling between matter and quintessence induces a
modification of the quintessence potential
\begin{equation}
V_{_{\rm eff}}(Q)=V_{_{\rm DE}}(Q)+A_{_{\rm CDM}}(Q)\frac{\rho _{_{\rm
      CDM}}^0}{a^3}\, ,
\end{equation}
where $A_{_{\rm CDM}}(Q)\equiv m_{_{\rm CDM}}(Q)/m_{_{\rm CDM}}(0)$,
$m_{_{\rm CDM}}$ being the mass of the dark matter particle, typically
the lightest supersymmetric particle. When the effective potential
admits a time-dependent minimum, the model is known as a chameleon
model~\cite{cham} and the cosmological evolution is changed. However,
in the present context, this correction is negligible in comparison
with the correction coming from the susy breaking, \ie $m_{3/2}^2Q^2$,
see Eq.~(\ref{newpotde}). In fact, one must compare the derivatives of
the two corrections since this is the derivative of the potential
which appears in the Klein-Gordon equation. For the new term coming
from the dark matter energy density, one has
\begin{equation}
\frac{\rho _{\rm mat}^0}{a^3}\frac{1}{A_{_{\rm CDM}}}\frac{\partial
  A_{_{\rm CDM}}}{\partial Q}A_{_{\rm CDM}}\sim \frac{\rho _{\rm
  mat}^0}{a^3}\kappa ^{1/2}\alpha _{\rm u,d}\sim \frac{\rho _{\rm
  mat}^0}{a^3}\kappa Q\, ,
\end{equation}
and for the susy breaking term, this is simply $m_{3/2}^2Q$. Since
$m_{3/2}^2\gg \rho _{\rm mat}/\mpl^2$, one can ignore the
correction coming from the cold dark matter energy density as
announced previously. This is essentially  due to the susy
breaking term as the field happens to settle down at a minimum
where its vev is very small. The situation would drastically
change if the potential kept its runaway shape as could be the
case, for instance, if the hidden sector were such that $a_i(Q)$
and $c_i(Q)$ were non trivial. In the following, we do not
consider this situation and, therefore, we neglect the correction
coming from the non trivial dependence of the dark matter energy
density.

\section{Implications for Cosmology}
\label{Implications for Cosmology}

\subsection{Fixing the Free Parameters of the Potential}
\label{Fixing the Free Parameters of the Potential}

In this section, we study the cosmological implications for the case
$a_i=c_i=0$ as we have just seen that this case cannot be ruled out by
the gravity constraints. The previous quintessence potential with
$a_i=c_i=0$ is characterized by three free parameters, $M$,
$m_{3/2}^0$ and $\Upsilon $. As a consequence, the potential can be
rewritten in terms of dimensionless quantities as
\begin{equation}
V_{\rm quint}(Q)=m^4 {\rm e}^{\bar{Q}^2/2}
\left(\bar{Q}^{-\alpha }+A\bar{Q}^2-B\right)\, ,
\end{equation}
where $\bar{Q}\equiv \kappa ^{1/2}Q$ is dimensionless and (obviously,
the $A$ and $B$ below have nothing to do with the soft terms
calculated in the previous subsection)
\begin{eqnarray}
\left(\frac{M}{\mP}\right)^{4+\alpha} &=& (8\pi
)^{-\alpha /2}\left(\frac{m}{\mP}\right)^4\, , \quad
A =\frac12 (8\pi )^{1-\alpha /2}\left(\frac{m_{3/2}^0}{\mP}\right)^2
\left(\frac{M}{\mP}\right)^{-(4+\alpha )}\, ,\\
B &=& (8\pi )^{-\alpha /2}\left(\frac{\Upsilon }{\mP}\right)^4
\left(\frac{M}{\mP}\right)^{-(4+\alpha )}\, .
\end{eqnarray}
The above potential has a minimum, $\bar{Q}_{\rm min}$. Let us assume
that we are in a situation where $\bar{Q}_{\rm min}\ll 1$. In this
case, an explicit expression of $\bar{Q}_{\rm min}$ can be
found. Since the value of the field at the minimum is small, the
exponential factor will be close to one. If so, it is easy to show
that
\begin{equation}
\bar{Q}_{\rm min}\simeq \left(\frac{p}{A}\right)^{1/(\alpha +2)}\, .
\end{equation}
We see that the location of the minimum is now controlled by the
gravitino mass through the constant $A$. It is interesting to compare
this value with the minimum of the usual SUGRA potential which reads
$\bar{Q}_{\rm min}=\sqrt{\alpha }$ and is controlled by the Planck
mass only. As we show in the following, the value of the minimum in
the case where supersymmetry breaking is taken into account is much
smaller than the minimum of the SUGRA potential. If we define the
reduced field $q$ by $q\equiv \bar{Q}/\bar{Q}_{\rm min}$, then the
potential becomes
\begin{equation}
V_{\rm quint}(q)=\tilde{m}^4{\rm e}^{\bar{Q}_{\rm
min}^2q^2/2}\left(q^{-\alpha }+pq^2-\tilde{B}\right)\, .
\end{equation}
The new scale $\tilde{m}$ is of course directly related to the scale
${m}$. The relation reads $\tilde{m}^4=\bar{Q}_{\rm min}^{-\alpha
}m^4$. The potential is still characterized by three parameters which
are now $\tilde{m}$, $\tilde{B}$ and $\bar{Q}_{\rm min}$. The constant
$\tilde{m}$ is chosen such that the quintessence energy density be
approximately $70\%$ of the critical energy density today. If we
assume that the field is, today, at its minimum, \ie~$q=1$, (this will
be shown in the following) this leads to
\begin{equation}
\tilde{m}^4\sim \frac{\Omega _Q}{1+\alpha /2-\tilde{B}}\rho _{\rm cri}
\simeq {\cal O}(1)\rho _{\rm cri}\, .
\end{equation}
Then, one can link the remaining parameters with the particle physics
parameters. For the gravitino mass, one obtains
\begin{equation}
\label{gravitinomass}
\left(\frac{m_{3/2}^0}{\mP}\right)^{2}\approx \frac{\alpha }{8\pi }
\bar{Q}_{\rm min}^{-2}\frac{\rho _{\rm cri}}{\mP^4}\, .
\end{equation}
As an example, one can take $m_{3/2}^0\simeq 100 \mbox{GeV}$ which
gives $\bar{Q}_{\rm min}\sim 2.6 \times 10^{-45}$, where we have used
$\mP\sim 1.22 \times 10^{19}\mbox{GeV}$ and $\rho _{\rm cri}\sim 8.1
h^2 10^{-47}\mbox{GeV}^4$ with $h\sim 0.72$ ($h$ is the reduced
present time Hubble parameter). If $m_{3/2}^0=1\mbox{eV}$, one obtains
$\bar{Q}_{\rm min}\simeq 2.6\times 10^{-34}$. Therefore, the minimum
is located at tiny values of the quintessence vev (compared to the
Planck mass). This makes a very important difference compared to the
standard SUGRA case where, as already mentioned before, the minimum is
close to $\mpl $. Notice that, despite the fact that the vev of the
quintessence field is now very small in comparison to the Planck
scale, this does not mean that supergravity is no longer a necessary
ingredient in the present context. This is because, when susy breaking
is taking into account, the global susy limit is more subtle than
simply taking the limit $\mpl \rightarrow +\infty $.

\par

Then, the scale $M$ is given by
\begin{equation}
\left(\frac{M}{\mP}\right)^{4+\alpha}=(8\pi )^{-\alpha /2}\bar{Q}_{\rm
min}^{\alpha }\left(\frac{\tilde{m}}{\mP}\right)^4\simeq (8\pi
)^{-\alpha /2}\bar{Q}_{\rm min}^{\alpha }\frac{\rho _{\rm
cri}}{\mP^4}\, .
\end{equation}
If one uses the expression of $\bar{Q}_{\rm min}$ deduced from
Eq.~(\ref{gravitinomass}) and the expression of the mass $M$,
$M^{4+\alpha }\sim \lambda ^2\xi ^4m_{\rm c}^{\alpha }$, one obtains
(assuming, as usual, $\lambda \sim 1$) the expression of the scale
$\xi $, namely
\begin{equation}
\frac{\xi }{\mpl}\sim \left(\frac{\rho _{\rm cri}}{\mpl }\right)^{1/4}
\left(\frac{\mpl}{m_{\rm c}}\right)^{\alpha /4}
\left[\left(\frac{m_{3/2}^0}{\mpl }\right)^{-2}\frac{\rho _{\rm
      cri}}{\mpl ^4}\right]^{\alpha /8}\, .
\end{equation}
This expression should be compared with Eq.~(\ref{xisugra}). We see
that the difference lies in the presence of the term in the squared
bracket which contains the gravitino mass and the critical energy
density. This has disastrous consequences. If, as before, one chooses
$m_{3/2}^0=100\mbox{GeV}$, then for, say, $\alpha =4$ one gets
$\xi/\mpl \sim 10^{-70}$. Even more serious, it is clear that there is
no value of $\alpha $ which, for a reasonable value of the gravitino
mass, would allow us to obtain a scale $\xi $ greater than the TeV as
it was the case for Eq.~(\ref{xisugra}). Recall that the energy scale
associated to the cosmological constant is $\Lambda /\mpl \sim
10^{-30}$. This means that, in the case of quintessence, we have to
build a complicated model and that, in addition, we have to fine tune
the basic scale of the model even more than in the case of the
cosmological constant where nothing else is needed. The origin of the
problem can really be traced back to the fact that modeling
quintessence in a more realistic fashion (breaking supersymmetry
properly, taking into account the interaction between the various
sectors etc ...) has modified the shape of the potential such that the
original success of finding reasonable $\xi $ from the ``see-saw''
formula~(\ref{xisugra}) has vanished.

\par

Finally, the scale $\Upsilon $ can be expressed as
\begin{equation}
\left(\frac{\Upsilon }{\mP}\right)^{4}\approx \tilde{B}\frac{\rho
_{\rm cri}}{\mP^4}\, ,
\end{equation}
and this implies that $\sum _{\alpha =1}^n\xi _{\alpha }^2\sim
3/\kappa $. This equation was used in Ref.~\cite{BMpart}.

\par

The previous estimates rest on the assumption that, very quickly, the
quintessence field is stabilized at its minimum. Considering to the
drastic consequences evoked before, we now carefully check that this
is indeed the case.

\subsection{Dynamics of the Quintessence Field}

Let us assume that the initial value of the field is such that
$Q_{\rm ini}\ll Q_{\rm min}$. In this case, the potential studied
above approximatively reduces to the Ratra-Peebles potential. In
fact, it seems necessary to start from the branch $Q^{-\alpha }$
rather from the branch $Q^2$ in order to have insensibility to the
initial conditions. Indeed, if the potential is made of a series
of monomial with positive powers,  it is known that a fine-tuning
of the initial conditions becomes necessary. In the case of the
Ratra-Peebles potential, there is a particular attractor solution
given by
\begin{equation}
\label{attra}
Q_{\rm attra}=Q_{\rm p}\left(\frac{a}{a_{\rm p}}\right)^{3(1+\omega
    _{_{\rm B}})/(\alpha +2)}\, ,
\end{equation}
where we have assumed that the field is a test field and that the
background behaves as
\begin{equation}
\label{scalefactor}
a(\eta )=a_{\rm p}\left(\frac{\eta }{\eta _{\rm
    p}}\right)^{2/(1+3\omega _{_{\rm B}})}\, ,
\end{equation}
$\eta $ being the conformal time and $\omega _{_{\rm B}}$ the equation
of state parameter of the background fluid (radiation or matter, that
is to say $\omega _{_{\rm B}}=1/3$ or $\omega _{_{\rm B}}=0$). The
time $\eta _{\rm p}$ is arbitrary and can be chosen at convenience. In
practice, we will often consider that this is the time of
reheating. The constant $Q_{\rm p}$ can be expressed as
\begin{eqnarray}
Q_{\rm p}^{-\alpha -2} &=& \frac{18}{\alpha ^2a_{\rm p}^2\eta _{\rm
p}^2M^{4+\alpha }} \frac{1-\omega _{\rm Q}^2}{(1+3\omega _{_{\rm
B}})^2}=\frac{9\left(1-\omega _Q^2\right)}{2\alpha ^2}\frac{H_{\rm
p}^2}{M^{4+\alpha }}\, .
\end{eqnarray}
where $H_{\rm p}$ denotes the Hubble parameter at time $\eta =\eta
_{\rm p}$. $\omega _Q$ is the equation of state parameter of the
attractor and is given by $\omega _Q=(-2+\alpha \omega _{_{\rm
B}})/(\alpha +2)$.

\par

At this point, it is worth recalling the following point, already
  discussed at the end of the previous section. For $\omega _{_{\rm
  B}}=0$, the solution~(\ref{scalefactor}) $a(\eta )\propto \eta ^2$
  is obtained from $\rho _{\rm mat}\propto 1/a^3$. However, in the
  present context, one has to be more precise as in Eq.~(\ref{rhomat})
  a dependence in the quintessence field is present. However, the $Q$
  dependence, at very small values of $Q$, is very weak. In other
  words $m_a\left(\kappa Q^2\right)\sim \mbox{cte}$ if $Q\ll
  \mpl$. Therefore, in the regime under consideration, one can safely
  assume that $\rho_{\rm mat}\propto 1/a^3$ and the corresponding
  behavior of the scale factor follows. In other words, as already
  discussed, one can safely neglect the fact that the model is a
  chameleon in the present context.

\par

The attractor solution is completely specified once the fact that
quintessence represents $70\%$ of the critical energy density today
has been imposed. Let us evaluate its value just after inflation. At
reheating, $z_{\rm reh}=10^{28}$, the value of the field is
\begin{equation}
\label{iniattra}
\bar{Q}_{\rm attra}\left(z_{\rm reh}\right)\simeq 10^{-117/(\alpha
+2)}\times \bar{Q}_{\rm min}^{\alpha /(\alpha +2)}\, ,
\end{equation}
where, for simplicity, we have not considered the factor $9(1-\omega
_Q^2)/(2\alpha ^2)$ which is, roughly speaking, of order one (in the
following we will always neglect this kind of factors since we are
interested in order of magnitude estimates only). We have also taken
the radiation contribution today to be $\Omega _{\rm rad}^0\sim
10^{-5}$. For $m_{3/2}^0=100\, \mbox{GeV}$ and $\alpha =6$,
$\bar{Q}_{\rm attra}(z_{\rm reh})\sim 8.6\times 10^{-49}\mP$. For
$m_{3/2}^0=1\mbox{eV}$ and the same parameters, $\bar{Q}_{\rm
attra}(z_{\rm reh})\sim 1.5 \times 10^{-40}\mP$. If we now compare the
initial quintessence energy density with the energy density of the
background (\ie~the energy density of radiation), one obtains
\begin{equation}
\Omega _{Q_{\rm attra}}\left(z_{\rm reh}\right)\sim
  10^{-107/(p+1)}\times \bar{Q}_{\rm min}^{2p/(p+1)}\, .
\end{equation}
For $m_{3/2}^0=100\, \mbox{GeV}$ and $\alpha =6$, one obtains
$\Omega_{Q_{\rm attra}}(z_{\rm reh})\sim 2.4\times 10^{-94}$. For
$m_{3/2}^0=1\mbox{eV}$ and $\alpha=6$, one obtains $\Omega_{Q_{\rm
attra}}(z_{\rm reh})\sim 7.4 \times 10^{-78}$. For comparison, we
recall that, at reheating, one has $\rho _{_{\rm B}}\sim 10^{107}\rho
_{\rm cri}$. The previous estimates show that the energy density of
the attractor is always initially very small in comparison with that
of the background. This is not the case in the ``standard''
Ratra-Peebles case. This difference is due to the fact that the scale
$M$ now depends on $\bar{Q}_{\rm min}$ which turns out to be a very
small quantity.

\par

The solution given in Eq.~(\ref{attra}) is valid as long as $Q_{\rm
attra}$ is small in comparison with $Q_{\rm min}$ and breaks down when
$Q_{\rm attra }\sim Q_{\rm min}$. This happens when
\begin{equation}
\frac{a_{\rm min}}{a_{\rm p}}\sim \left(\frac{H_{\rm
    p}}{m_{3/2}^0}\right)^{2/(3+3\omega _{_{\rm B}})}\, ,
\end{equation}
that is to say at a redshift given approximately by (strictly
speaking, this expression is valid if $z_{\rm min}> 10^4$ since we
used that $\omega _{_{\rm B}}=1/3$; otherwise, one has to take into
account that the background becomes matter dominated)
\begin{equation}
1+z_{\rm min} \sim 10^{32} \times
\left(\frac{m_{3/2}^0}{\mP}\right)^{1/2}\sim 10\times \bar{Q}_{\rm
min}^{-1/2}\, .
\end{equation}
For our typical examples with $m_{3/2}^0\simeq 100 \, \mbox{GeV}$, one
obtains $z_{\rm min}\simeq 2.4 \times 10^{23}$ to be compared with the
reheating redshift, $z_{\rm reh}\sim 10^{28}$. For
$m_{3/2}^0=1\mbox{eV}$, one gets $z_{\rm min}\simeq 7.7 \times
10^{17}$. In both cases, the field reaches the minimum well before Big
Bang Nucleosynthesis.

\par

Now, the field is of course not necessarily on the attractor
initially. It is therefore important to estimate at which redshift the
attractor is joined and to compare this redshift to $z_{\rm min}$. If
$Q_{\rm ini}>Q_{\rm attra}$ (undershoot) then the field remains
frozen. Therefore, the redshift at which the attractor is joined is
given by the condition $Q_{\rm ini}=Q_{\rm attra}$ which results in
\begin{equation}
\frac{a_{\rm under}}{a_{\rm p}}\sim \left(Q_{\rm
ini}^{2p+2}\frac{H_{\rm p}^2}{M^{4+2p }}\right)^{1/(3+3\omega _{_{\rm
B}})}\, .
\end{equation}
Using this relation, one finds
\begin{equation}
1+z_{\rm under}\sim 10\times \left(\frac{\bar{Q}_{\rm
    min}}{\bar{Q}_{\rm ini}}\right)^{p/2} \bar{Q}_{\rm ini}^{-1/2}\, .
\end{equation}
One can check that, if the initial value of the field is the value on
the attractor given by Eq.~(\ref{iniattra}), then $z_{\rm under}\sim
z_{\rm reh}$ as expected. More generally, we conclude that each time
the field starts in a undershoot situation, that is to say initially
$Q_{\rm attra}<Q_{\rm ini}<Q_{\rm min}$, it will join the attractor
before reaching the minimum.

\par

Let us now consider the case of an overshoot, \ie~$Q_{\rm ini}<Q_{\rm
attra}$. Then, the field is first kinetic dominated until the
potential energy becomes equal to the kinetic energy. When this
happens, the field becomes frozen until the attractor is
joined. During the phase dominated by the kinetic energy, the field
behaves as
\begin{equation}
\label{Qkinetic}
Q=Q_{\rm ini}+\mP \sqrt{\frac{3\Omega _{\rm Q_{\rm ini}}}{4\pi }}
\left(1-\frac{a_{\rm ini}}{a}\right)\, .
\end{equation}
This allows us to estimate the redshift at which the field becomes
frozen, $z_{\rm froz}$, and to compare it with $z_{\rm min}$. The
redshift at which the field becomes frozen can be estimated to be
\begin{equation}
1+z_{\rm froz}\sim 10^{10}\times \bar{Q}_{\rm min}^{p/3}\times
\Omega _{Q_{\rm ini}}^{-(p+1)/6}\, .
\end{equation}
On the other hand, if the quintessence field behaves as in
Eq.~(\ref{Qkinetic}), it reaches the minimum at a redshift given by
\begin{equation}
1+z_{\rm kin\rightarrow min}\sim 10^{10}\times \Omega _{Q_{\rm ini}}^{-1/6}\, .
\end{equation}
Therefore, the field becomes frozen before reaching the minimum if
$1+z_{\rm froz}>1+z_{\rm kin\rightarrow min}$, a condition which
amounts to $\bar{Q}_{\rm min}>\Omega _{\rm Q_{\rm ini}}^{1/2}$. As an
illustration, let us consider again the model for which
$m_{3/2}^0=100\, \mbox{GeV}$ and $\alpha =6$. From the previous
considerations, one knows that $\bar{Q}_{\rm min}\simeq 2.6 \times
10^{-45}$ and that, in order to have overshoot, $10^{-94}\lta \Omega
_{Q_{\rm ini}}\lta 10^{-4}$, this last bound corresponding to
equipartition initially. Therefore, for the above condition to be
satisfied, one needs $ \Omega _{Q_{\rm ini}}\lta 10^{-90}$.

\par

We conclude that, on more general grounds, the above condition cannot
be satisfied unless the field is, at the beginning, very close to the
attractor. Therefore, generically, when we have overshoot, the field
reaches the minimum in the kinetic dominated phase and has no time to
freeze out.

\subsection{Approaching the Minimum}

We now describe the behavior of the quintessence field when it starts
feeling that the potential has developed a minimum. When the field is
close to the minimum, the potential can be approximated by
\begin{equation}
V(Q)\simeq V_{\rm min}+\frac12 \left(m_{3/2}^0\right)^2\left(Q-Q_{\rm
min}\right)^2\, .
\end{equation}
The mass of the field is the gravitino mass as established before and
in Ref.~\cite{BMpart}. Let us first consider the situation where the
quintessence field is a test field. The Klein-Gordon equation, written
with the number of e-folds $N$ as the time variable, reads
\begin{eqnarray}
\frac{{\rm d}^2}{{\rm d}N^2}\left(\frac{Q}{\mP}\right) &
&+\left(3+\frac{1}{H}\frac{{\rm d}H}{{\rm d}N}\right) \frac{{\rm
d}}{{\rm d}N}\left(\frac{Q}{\mP}\right)+\frac{\mP
^2}{H^2}\frac{\partial f}{\partial (Q/\mP)}=0\, ,
\end{eqnarray}
where $V(Q)\equiv \mP^4f(Q)$. The Hubble parameter is given by
$H=H_{\rm p}\exp[-3(1+\omega _{_{\rm B}})N_{\rm p}/2]$, where $N_{\rm
p}$ is the number of e-folds counted from the time $\eta =\eta _{\rm
p}$. The above equation can be easily integrated and the solution
reads
\begin{eqnarray}
\label{solmin}
\frac{Q}{\mP} &=& \frac{Q_{\rm min}}{\mP}+{\rm e}^{-3(1-\omega _{_{\rm
    B}})N_{\rm p}/4} \biggl\{A_1J_{(1-\omega _{_{\rm B}})/(2+2\omega
    _{_{\rm B}})} \left[\frac{2}{3(1+\omega _{_{\rm
    B}})}\frac{m_{3/2}^0}{H_{\rm p}} {\rm e}^{3(1+\omega _{_{\rm
    B}})N_{\rm p}/2}\right] \nonumber \\ & & +A_2J_{-(1-\omega _{_{\rm
    B}})/(2+2\omega _{_{\rm B}})} \left[\frac{2}{3(1+\omega _{_{\rm
    B}})}\frac{m_{3/2}^0}{H_{\rm p}} {\rm e}^{3(1+\omega _{_{\rm
    B}})N_{\rm p}/2}\right]\biggr\}\, ,
\end{eqnarray}
where $J_{\nu }(z)$ is a Bessel function of order $\nu $. As expected
the field start oscillating around its minimum when its mass equals
the Hubble parameter. Using the expression of the gravitino mass given
before, one easily checks that this happens at a redshift of
\begin{equation}
1+z_{\rm osci}\sim 10\times \left(\bar{Q}_{\rm min}\right)^{-1/2}.
\end{equation}
When the mass is smaller than the Hubble parameter, the field is
essentially frozen. Using the asymptotic expression of the Bessel
functions for small arguments, one obtains
\begin{equation}
\label{solminbefore}
\frac{Q}{\mP} \simeq \frac{Q_{\rm min}}{\mP}+\bar{A}_1 +\bar{A}_2{\rm
e}^{-3(1-\omega _{_{\rm B}})N_{\rm p}/2}\, ,
\end{equation}
where $\bar{A}_1$ and $\bar{A}_2$ are two new constants, different
from $A_1$ and $A_2$. Very rapidly, the branch proportional to
$\bar{A}_2$ becomes negligible. In the opposite situation, \ie~when
the mass is much larger than the Hubble parameter, one can use the
asymptotic expansion of the Bessel functions for large values of the
arguments and one obtains
\begin{eqnarray}
\label{solminafter} \frac{Q}{\mP} &\simeq & \frac{Q_{\rm
min}}{\mP}+{\rm e}^{-3(1-\omega _{_{\rm B}})N_{\rm
p}/2}\Biggl\{\tilde{A_1} \cos \left[\frac{2}{3(1+\omega _{_{\rm
B}})}\frac{m_{3/2}^0}{H_{\rm p}} {\rm e}^{3(1+\omega _{_{\rm
B}})N_{\rm p}/2}-\frac{\pi (1-\omega _{_{\rm B}})}{4(1+\omega _{_{\rm
B}})}-\frac{\pi }{4}\right] \nonumber \\ & & +\tilde{A_2} \cos
\left[\frac{2}{3(1+\omega _{_{\rm B}})}\frac{m_{3/2}^0}{H_{\rm p}}
{\rm e}^{3(1+\omega _{_{\rm B}})N_{\rm p}/2}+\frac{\pi (1-\omega
_{_{\rm B}})}{4(1+\omega _{_{\rm B}})}-\frac{\pi }{4}\right]\Biggr\}\,
,
\end{eqnarray}
where $\tilde{A}_1$ and $\tilde{A}_2$ are new constants. The
oscillations are damped by a factor $a^{-3(1-\omega _{_{\rm B}})/2}$.

\par

Let us summarize the two possibilities. In case of an undershoot, the
field joins the attractor and then reaches the minimum without any
oscillatory phase as $z_{\rm osci}\sim z_{\rm min}$. If there is an
overshoot, the field has no time to freeze out and goes directly from
the kinetic dominated phase to the oscillatory phase.

\par

At some point, the quintessence field starts dominating the matter
content of the Universe. In this case, the above treatment breaks down
since the quintessence field is no longer a test field. Assuming, for
simplicity, that $\rho \sim V_{\rm min}$, the Klein-Gordon equation
can still be solved explicitly. The solution reads
\begin{eqnarray}
\frac{Q}{\mP}&=&\frac{Q_{\rm min}}{\mP}\nonumber +{\rm
e}^{-3N/2} \Biggl\{B_1\cos \left[\sqrt{\frac{3}{8\pi
}}\frac{m_{3/2}^0/\mP}{(V_{\rm min}/\mP^4)^{1/2}}N\right]
\nonumber \\ & &
+B_2\sin\left[\sqrt{\frac{3}{8\pi }}\frac{m_{3/2}^0/\mP}{(V_{\rm
min}/\mP^4)^{1/2}}N\right]\Biggr\}\,
\end{eqnarray}
and we still have very rapid oscillations, damped by a factor
$a^{-3/2}$. For $m_{3/2}^0=100 \, \mbox{GeV}$, the dimensionless
frequency ($N$ being the time variable) is $\sim 10^{43}$. Again the
oscillations stop rapidly as their amplitude decays exponentially with
the number of e--folds.

\subsection{Numerical Integration}

Of course, rather than the approximate considerations developed above,
a full numerical integration would allow us to obtain the exact
solution. Unfortunately, the realistic values of $\bar{Q}_{\rm min}$
are so small that a simple Fortran code cannot handle the
corresponding solution. However, one can check the previous analytical
estimates for values of $\bar{Q}_{\rm min}$ which are numerically
reasonable (but not physically realistic). Then, having checked and
validated the previous estimates, we will use them in a physically
relevant situation.

\par

Let us consider a situation where the free parameters of the potential
are given by $\bar{Q}_{\rm min}=10^{-4}$, $\alpha =6$ and
$\tilde{B}=1$. This implies that the mass scale $M\sim 8 \times
10^{-16}\mP$, the gravitino mass $m_{3/2}^0\sim 2 \times 10^{-58}\mP$
and $\Upsilon \sim 2 \times 10^{-31}\mP$. According to the previous
estimates, the initial value of the field on the attractor is
$\bar{Q}_{\rm attra}\sim 2.3 \times 10^{-18}\mP$ corresponding to
$\Omega _{Q_{\rm attra}}\sim 1.8 \times 10^{-33}$. On the attractor,
the minimum of the potential is felt at a redshift of $1+z_{\rm
min}\sim 1000$.

\par

Let us now consider the initial conditions corresponding to
equipartition, \ie $\Omega _{Q_{\rm ini}}=10^{-4}$. This implies that
$\bar{Q}_{\rm ini}\sim 6.8\times 10^{-22}\mP$ and we have overshoot
since $\bar{Q}_{\rm ini}<\bar{Q}_{\rm attra}$. As a consequence, as
explained before, the initial evolution is dominated by the kinetic
energy and we have
\begin{equation}
\label{Qkinetic2}
Q=Q_{\rm ini}+\mP \sqrt{\frac{3\Omega _{\rm Q_{\rm ini}}}{4\pi }}
\left(1-{\rm e}^{-N}\right)\, ,
\end{equation}
where $N$ is the total number of e-folds counted from reheating. Very
quickly, we have $Q=\sqrt{3\Omega _{\rm Q_{\rm ini}}/(4\pi )}\sim
0.00489 \mP$ (or $\bar{Q}\sim 0.0245 \mP$). Let us notice that this
value is greater than the value of the minimum. Although the field
seems to be frozen, its time variation is still sufficient for the
corresponding kinetic energy to be greater than the critical energy (a
similar situation arises in the standard Ratra-Peebles scenario, see
Ref.~\cite{BMR1}). In fact, the field rolls down the potential so
quickly that it goes through the minimum while the kinetic regime goes
on (when the kinetic energy dominates, the fact that the potential has
a minimum is irrelevant). The kinetic energy reaches the critical
energy, by definition at $z_{\rm kin\rightarrow min}$, when the field
is on the other side of the potential.  As a consequence, the field
will approach the minimum ``from the right''.

\par

In the Ratra-Peebles case, the kinetic regime comes to an end at the
  redshift $1+z_{\rm froz}\sim 4.64\times 10^8$ while, in the present
  case where the potential possesses a minimum, this one is felt by
  the field at $z_{\rm kin \rightarrow min}\sim 4.64 \times
  10^{10}$. Therefore, in this case, the field does not enter the
  potential dominated regime at all and directly goes from the kinetic
  dominated regime to a regime where the minimum is felt and where the
  solution~(\ref{solmin}) is relevant. However, the solutions before
  $z_{\rm kin \rightarrow min}\sim 4.64 \times 10^{10}$ given by
  Eq.~(\ref{Qkinetic2}) and after, given by Eq.~(\ref{solminbefore})
  are the same. As a consequence, the coefficients $\bar{A}_1$ and
  $\bar{A}_2$ are such that the solution after $z_{\rm kin \rightarrow
  min}$ is still given by Eq.~(\ref{Qkinetic2}). So, even after
  $z_{\rm kin \rightarrow min}$, the field remains ``frozen'' at
  $Q\sim 0.00489 \mP$. Nevertheless, the evolution of the energy
  density changes and, instead of $\rho \propto a^{-6}$, we have $\rho
  \sim \mbox{cte}$.

\par

\begin{figure*}[t]
  \includegraphics[width=14cm]{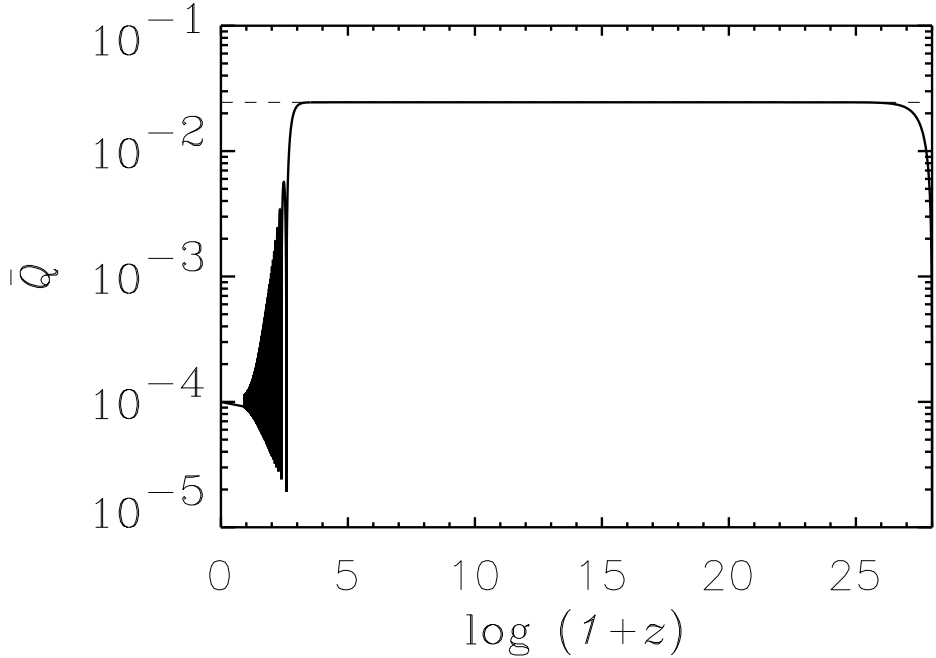}
  \caption{Evolution of the scalar field for the parameters,
  $\bar{Q}_{\rm min}=10^{-4}$, $\tilde{B}=1$ and $\alpha =6$ with the
  initial conditions such that $\Omega _{Q_{\rm ini}}=10^{-4}$
  (equipartition and overshoot). The dashed horizontal represents the
  value $Q=\sqrt{3\Omega _{Q_{\rm ini}}/(4 \pi )}$.  It is clear from
  this plot that the numerical estimates presented in the text are
  fully compatible with the numerical evolution. In particular, one
  can check that the oscillations start at $z\sim 10^{3}$ and that,
  eventually, after rapid oscillations, the field stabilizes at its
  minimum.}
\label{overshootfield}
\end{figure*}

\begin{figure*}[t]
  \includegraphics[width=14cm]{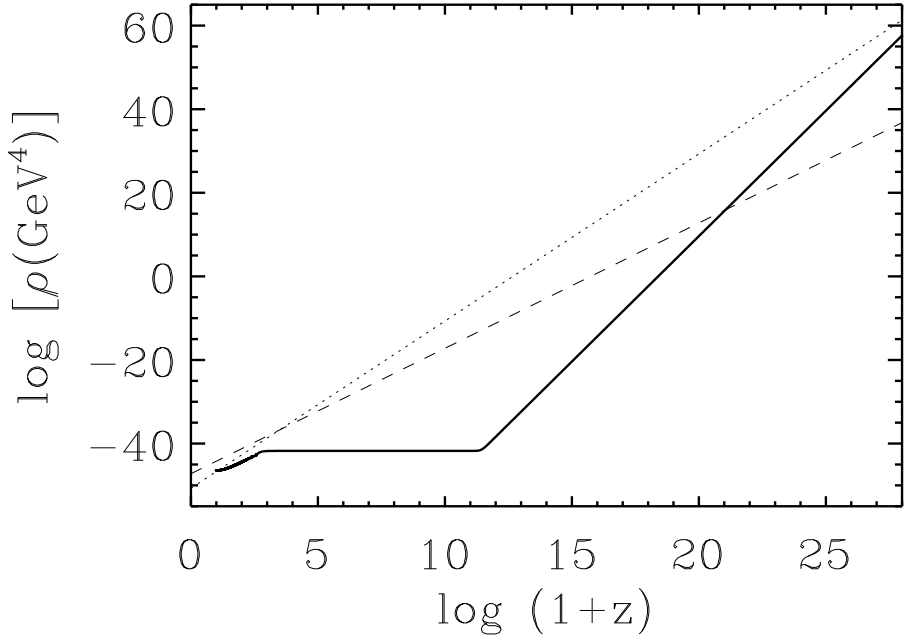}
  \caption{Evolution of the quintessence energy density (solid line)
  for $\bar{Q}_{\rm min}=10^{-4}$, $\tilde{B}=1$ and $\alpha =6$ with
  the initial conditions such that $\Omega _{Q_{\rm
  ini}}=10^{-4}$. The dotted line represents the evolution of the
  radiation energy density while the dashed line is the matter energy
  density. The energy density freezes at $z_{\rm kin \rightarrow min}$
  and not at $z_{\rm froz}$ as predicted in the text since the minimum
  of the potential is felt before the kinetic energy becomes equal to
  the potential energy. The analytical estimate $z_{\rm kin
  \rightarrow min}\sim 5 \times 10^{10}$ is in relatively good
  agreement with the actual value observed in the figure. The
  difference (about one order of magnitude) is probably due to the
  fact that the minimum is felt before the energy density is equal to
  $\rho _{\rm cri}$ (which was the criterion used in order to derive
  the expression of $z_{\rm kin \rightarrow min}$) which has the
  effect to increase $z_{\rm kin \rightarrow min}$.}
\label{overshootenergy}
\end{figure*}

At $1+z_{\rm osci}\sim 10^3$, the Hubble parameter is equal to the
gravitino mass and the damped oscillations of Eq.~(\ref{solminafter})
start. Then, the field stabilizes at its minimum. The evolution of the
field is represented in Fig.~\ref{overshootfield} and the
corresponding quintessence energy density is plotted in
Fig.~\ref{overshootenergy}.

\begin{figure*}[t]
  \includegraphics[width=14cm]{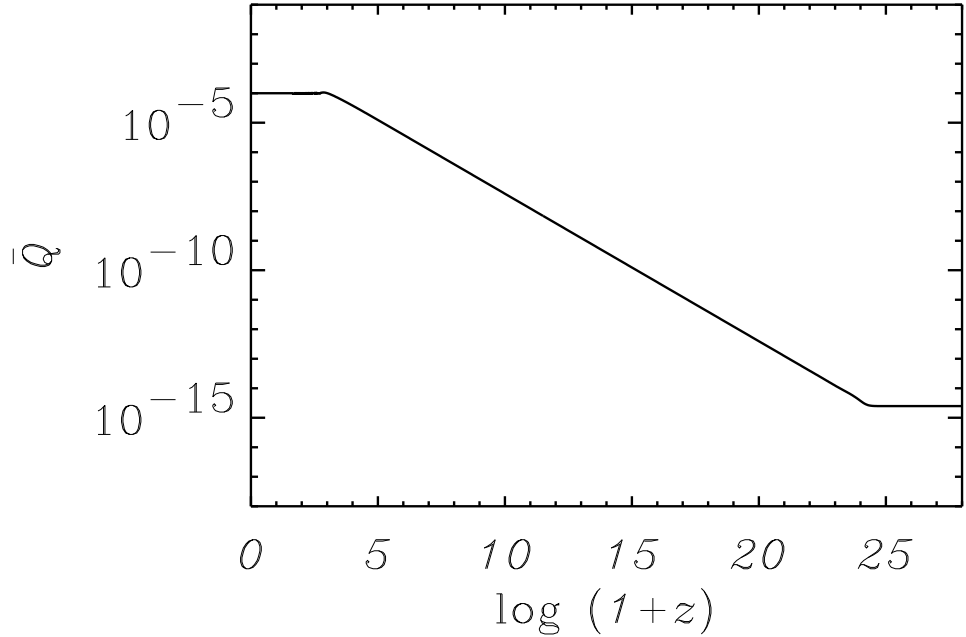}
  \caption{Evolution of the scalar field for the parameters,
  $\bar{Q}_{\rm min}=10^{-4}$, $\tilde{B}=1$ and $\alpha =6$ with the
  initial conditions such that $\Omega _{Q_{\rm ini}}=10^{-45}$
  (undershoot). The numerical estimates presented in the text are in
  good agreement with the numerical evolution.}
\label{undershootfield}
\end{figure*}

\begin{figure*}[t]
  \includegraphics[width=14cm]{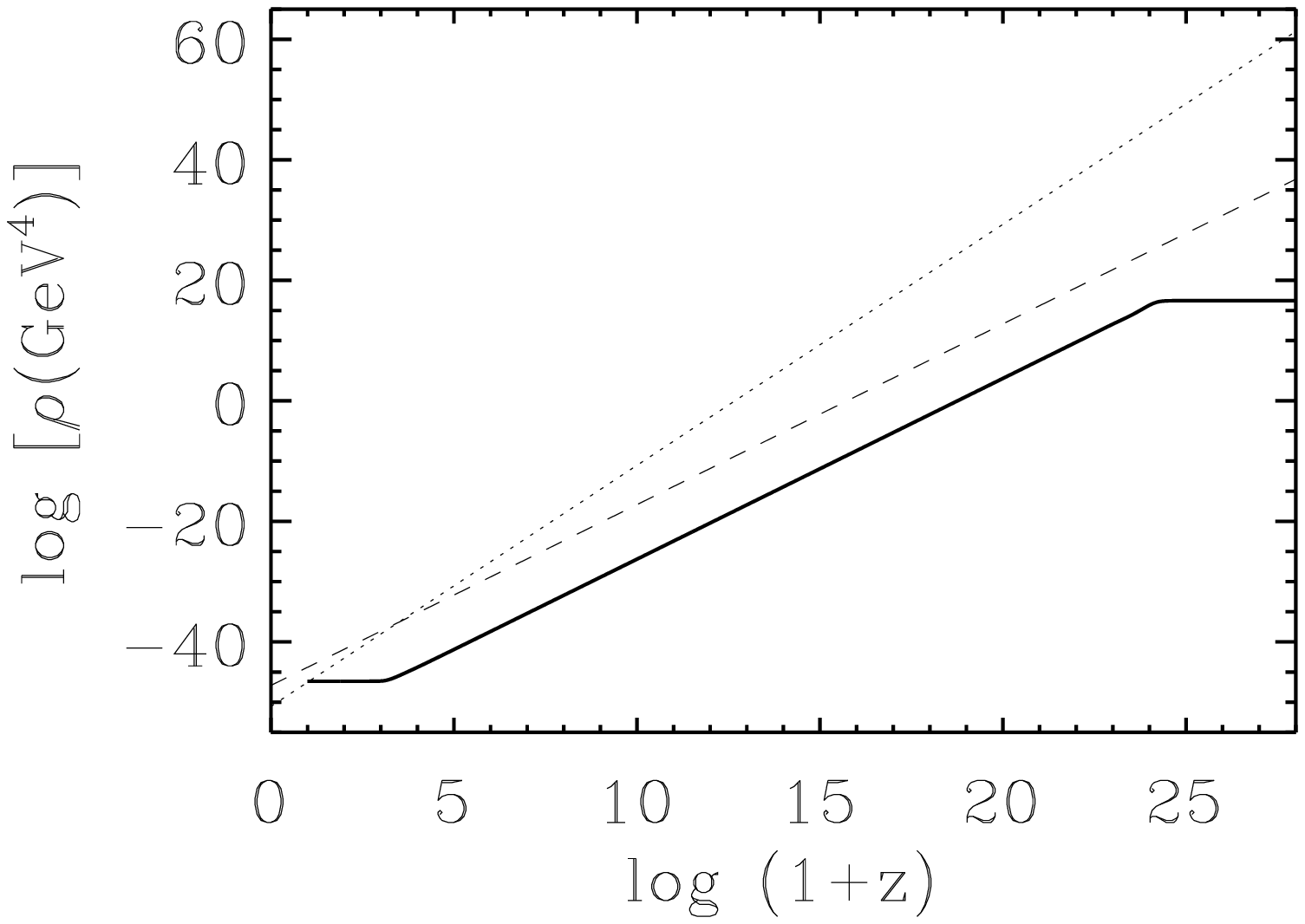}
  \caption{ Evolution of the quintessence energy density (solid line)
  for $\bar{Q}_{\rm min}=10^{-4}$, $\tilde{B}=1$ and $\alpha =6$ with
  the initial conditions such that $\Omega _{Q_{\rm
  ini}}=10^{-45}$. The dotted line represents the evolution of the
  radiation energy density while the dashed line is the matter energy
  density. Initially, the energy density is frozen until the attractor
  is joined at $z_{\rm under}\sim 4.6 \times 10^{23}$. At $z_{\rm
  min}=z_{\rm osci}$, the presence of the minimum is felt by the field
  which stabilizes at its minimum.}
\label{undershootenergy}
\end{figure*}

Let us now consider the case of an undershoot, \ie~the case where the
initial value of the field is greater than the initial value on the
attractor. We take $\Omega _{Q_{\rm ini}}=10^{-45}$ which implies that
$Q_{\rm ini}\sim 4.6 \times 10^{-15}\mP$. The corresponding numerical
evolution is represented in Figs.~\ref{undershootfield}
and~\ref{undershootenergy}. Since we have undershoot, the field will
be initially frozen until it reaches the attractor. This happens at
$z_{\rm under}\sim 4.6 \times 10^{23}$. On the attractor, the
effective equation of state is $\omega _Q=(-2+\alpha \omega _{_{\rm
B}})/(\alpha +2)=0$ for $\alpha =6$ and $\omega _{_{\rm B}}=1/3$. As a
consequence, the energy density of quintessence scales as matter, as
can be checked on Fig.~\ref{undershootenergy}, until the minimum is
reached. The minimum is felt at $z_{\rm min}\sim 1000$ which is also
$z_{\rm osci}$, the redshift at which the Hubble parameter is equal to
the gravitino mass and the oscillations start. Therefore, we expect no
phase where the field is frozen, as predicted by
Eq.~(\ref{solminbefore}) but expect the oscillations to start
immediately after the field has left the attractor. However, in the
case of an undershoot, when the minimum is felt we necessarily have
$Q\sim Q_{\rm min}$. In the previous case of an overshoot this was not
the case because the kinetic energy of the field was dominating the
potential energy at the moment where the presence of the minimum is
seen by the field. Therefore, the amplitude of the oscillations is
very small and, in practice, we see no oscillations at all. The field
directly stabilizes at this minimum. This behavior is indeed observed
in Fig.~\ref{undershootfield}.

As a conclusion of the subsection, we have checked that the numerical
estimates derived before are confirmed by a numerical calculation in
the (physically unrealistic) case where this one is possible.

\par

Let us summarize our findings in the realistic cases where the
gravitino mass is of the eV or 100 GeV scales. In both cases the
minimum of the potential is at extremely small values compared to the
Planck scale. This implies that we can safely neglect the $Q$
dependence of the particle masses and trust the supergravity expansion
scheme in $1/m_{\rm c}$. On the other hand, the existence of a minimum
with a mass of the order of the gravitino mass leads to a drastic
modification of the evolution of the quintessence field.  In both
cases, either overshoot or undershoot, we find that the quintessence
field settles down at the bottom of the potential before the beginning
of BBN. From the point of view of late time physics, the quintessence
sector plays the role of an effective cosmological constant with
equation of state $\omega _Q=-1$. Observationally this is not a
problem since a cosmological constant is perfectly compatible with the
currently available data. However, conceptually, we consider it as a
second fundamental difficulty (the first one was he magnitude of $\xi
$) of the realistic model where supersymmetry breaking is taken into
account. Indeed, it seems clear that the justification for building a
complicated model which is just equivalent to a cosmological constant
is very weak. One possible way out is to study whether the behavior of
the perturbed quantities allows us to distinguish this model from a
pure cosmological constant. In particular, although the equation of
state is $\omega _Q=-1$, the mass of the quintessence field is now
$m_{3/2}^0$ and, therefore, the corresponding Jeans length is very
small compared to the usual case where it is the Hubble scale. This is
why, in the next section, we analyze whether the perturbations of the
quintessence field can give information on the dynamics of
quintessence prior to BBN.

\subsection{Cosmological Perturbations}
\label{Cosmological Perturbations}

In this section, we study how the perturbations of the quintessence
field behave. In conformal time, the perturbed Klein-Gordon equation
Fourier space reads
\begin{equation}
\delta Q_{\bf k}'' +2{\cal H}\delta Q_{\bf k}'+\left(k^2+a^2\frac{{\rm
  d}^2V}{{\rm d}Q^2}\right) \delta Q_{\bf k}=4Q'\Phi _{\bf
  k}'-2a^2\frac{{\rm d}V}{{\rm d}Q}\Phi _{\bf k}\, ,
\end{equation}
where ${\cal H}\equiv a'/a$ and $\Phi $ is the Bardeen potential which
describes the metric perturbations. We want to study the situation
where the background field stands at its non-vanishing minimum, this
field being a test field, \ie~the scale factor being given by the
expression~(\ref{scalefactor}). In this case, the right hand side of
the above expression vanishes and the perturbed Klein-Gordon equation
reduces to
\begin{equation}
\label{pertkg}
\delta Q_{\bf k}'' +\frac{4}{\left(1+3\omega _{_{\rm
  B}}\right)\eta}\delta Q_{\bf
  k}'+\left[k^2+\left(m_{3/2}^0\right)^2a_{\rm p}^2\left(\frac{\eta
  }{\eta _{\rm p}}\right)^{\frac{4}{1+3\omega _{_{\rm
  B}}}}\right]\delta Q_{\bf k}=0\, .
\end{equation}
It is clear that the Jeans mass of the field is now
$(m_{3/2}^0)^{-1}$. Since this scale is much smaller than the Hubble
length, \ie~the Jeans length in the standard Ratra-Peebles case, this
raises the question as whether quintessence could collapse and develop
structures at very small scales. Unfortunately, the above equation
cannot be solved explicitly. However, it can be analyzed in the case
where the wavelength of the fluctuation is either larger or smaller
than the Jeans length.

\par

Let us start by assuming that $k_{\rm ph}\gg m_{3/2}^0$, \ie~the
wavelength of the Fourier mode is smaller than the Jeans length.
Then the term proportional to $(m_{3/2}^0)^2a^2$ in the perturbed
Klein-Gordon equation can be neglected. In this case, the solution
reads
\begin{equation}
\delta Q_{\bf k}=A_1(k)\eta ^{\nu }{\rm J}_{\nu }(k\eta )+A_2(k)\eta
^{\nu}{\rm J}_{-\nu }(k\eta )\, ,
\end{equation}
where $A_1(k)$ and $A_2(k)$ are two integration constants fixed by the
initial conditions which are not important for us in the present
context and where ${\rm J}_{\nu }$ is a Bessel function of order $\nu
$ which is a function of the background equation of state only
\begin{equation}
\nu =\frac{3(\omega _{_{\rm B}}-1)}{2(1+3\omega _{_{\rm
      B}})}\, .
\end{equation}
Moreover, $k_{\rm ph}\gg m_{3/2}^0$ implies $k_{\rm ph}\gg H$ and
then using the asymptotic behavior of Bessel functions for large
values of their argument, one obtains the approximate expression
\begin{eqnarray}
\delta Q_{\bf k} &\sim& \sqrt{\frac{2}{\pi }}\eta ^{\nu
  -1/2}\biggl[A_1(k)\cos\left(k\eta -\frac{\pi \nu}{2}-\frac{\pi
  }{4}\right) +A_2(k)\cos\left(k\eta +\frac{\pi \nu}{2}-\frac{\pi
  }{4}\right)\biggr]\, .
\end{eqnarray}
We have oscillations since we consider modes of wavelength of smaller
than the Jeans length. The overall amplitude behaves as $\eta
^{\nu-1/2}$. Since one has
\begin{equation}
\nu -\frac12=-\frac{2}{1+3\omega _{_{\rm B}}}<0\, ,
\end{equation}
there is no growing mode in this situation. This result makes sense as
the pressure forces counter balance the gravitational force which
tends to make the system collapse.

\par

Let us now consider the situation where $k_{\rm ph}\ll m_{3/2}^0$,
\ie~where the wavelength of the Fourier mode under consideration is
larger than the Jeans length $(m_{3/2}^0)^{-1}$.  In this case,
ignoring the term $k^2$ in Eq.~(\ref{pertkg}), the equation for
$\delta Q_{\bf k}$ can also be integrated exactly. It is interesting
to compare this equation with the standard equation (\ie~in the case
where the potential is the Ratra-Peebles one, $V=M^{4+\alpha
}Q^{-\alpha }$) for $\delta Q_{\bf k}$ on large scales when the field
is on the attractor. This case was studied in Ref.~\cite{BMR1}. On the
attractor, the second derivative of the potential is given by ${\rm
d}^2V/{\rm d}Q^2=9H^2(\alpha +1)(1-\omega _Q^2)/(2\alpha )$, $w_Q$
being a constant, and, as a result, the terms $a^2{\rm d}^2V/{\rm
d}Q^2$ scales as $\eta ^{-2}$. This is why the homogeneous equation
admits simple power-law solutions. Here, the term ${\rm d}^2V/{\rm
d}Q^2$ is a constant equal to $m_{3/2}^0$ and, therefore the time
dependence of the term $a^2{\rm d}^2V/{\rm d}Q^2$ is now given by
$a^2$, \ie~a power-law of the conformal time, namely $\eta ^2$ for the
radiation dominated era and $\eta ^4$ for the matter dominated
epoch. Therefore, we expect the solutions for $\delta Q_{\bf k}$ on
large scales to be different from the Ratra-Peebles case. Indeed, the
result reads
\begin{eqnarray}
\delta Q_{\bf k} &=& B_1(k)\eta ^{\frac{3(\omega _{_{\rm
    B}}-1)}{1+3\omega _{_{\rm B}}}}{\rm J}_{\mu} \left[\frac{1+3\omega
    _{_{\rm B}}}{3(1+\omega _{_{\rm B}})}m_{3/2}^0\eta _{\rm
    p}^{\frac{-2}{1+3\omega _{_{\rm B}}}} \eta ^{\frac{3(1+\omega
    _{_{\rm B}})}{1+3\omega _{_{\rm B}}}}\right] \nonumber \\ & & +
    B_2(k)\eta ^{\frac{3(\omega _{_{\rm B}}-1)}{1+3\omega _{_{\rm
    B}}}}{\rm J}_{-\mu} \left[\frac{1+3\omega _{_{\rm B}}}{3(1+\omega
    _{_{\rm B}})}m_{3/2}^0\eta _{\rm p}^{\frac{-2}{1+3\omega _{_{\rm
    B}}}} \eta ^{\frac{3(1+\omega _{_{\rm B}})}{1+3\omega _{_{\rm
    B}}}}\right]\, ,
\end{eqnarray}
where $B_1(k)$ and $B_2(k)$ are two arbitrary constants and where the
order $\mu $ of the Bessel functions can be written as
\begin{equation}
\mu =-\frac{\omega _{_{\rm B}}-1}{\omega _{_{\rm B}}+1}\, .
\end{equation}
The same expression can be also rewritten in a form for which the
physical interpretation is easier
\begin{eqnarray}
\delta Q_{\bf k} &=& B_1(k)\eta ^{\frac{3(\omega _{_{\rm
    B}}-1)}{1+3\omega _{_{\rm B}}}}{\rm J}_{\mu}
    \left[\frac{2}{3(1+\omega _{_{\rm B}})}\frac{m_{3/2}^0}{H}\right]
    + B_2(k)\eta ^{\frac{3(\omega _{_{\rm B}}-1)}{1+3\omega _{_{\rm
    B}}}}{\rm J}_{-\mu}\left[\frac{2}{3(1+\omega _{_{\rm
    B}})}\frac{m_{3/2}^0}{H}\right] \, . \nonumber \\ & &
\end{eqnarray}
 where $m_{3/2}^0\gg H$. Therefore, in the above equation, one can use
the asymptotic expression of the Bessel functions for large values of
their argument. This means that the overall amplitude behave as $\eta
^{3(\omega _{_{\rm B}}-1)/(1+3\omega _{_{\rm B}})-1/2}$. This gives
$\eta ^{-7/2}$ and $\eta ^{-3/2}$ for $\omega _{_{\rm B}}=0,
1/3$. Again this function is a decreasing function of $\eta$. Hence no
growing mode is generated by perturbation of the quintessence scalar
field.

\par

Together with the fact that for most of the history of the Universe,
the quintessence field sits at the minimum of its potential, this
implies that the dynamics of quintessence when coupling to
supersymmetry breaking in a hidden sector leads to an absence of
deviations from a pure cosmological constant in the late Universe,
even at the perturbed level and even though the Jeans length of the
field is now considerably smaller than in the standard Ratra-Peebles
case.

\section{Conclusions}
\label{Conclusions}

We have presented a cosmological analysis of quintessence models in
supergravity coupled to matter. For the quintessence sector, we have
used the SUGRA model whose main feature is to reduce to the
Ratra--Peebles potential at small values of the quintessence field.
This is effectively the only property which is required here. Hence
our results generalize to quintessence models with the same type of
potentials. Supersymmetry is broken in a hidden sector which couples
gravitationally both to the observable sector, \ie~the MSSM, and the
quintessence sector. Requiring that the gravitino mass is large enough
to lead to acceptable masses for the sparticles implies that the
quintessence potential is drastically modified by the presence of the
hidden sector. In particular, as soon as the hidden sector fields are
stabilized and supersymmetry is broken, we find that the quintessence
potential develops a minimum acting as an attractor for the
quintessence field in the very early universe. Indeed the quintessence
field settles down at the minimum of the potential before BBN and both
at the background and perturbation level, the model becomes equivalent
to a pure cosmological constant scenario. This is the first
quintessential difficulty as no observational consequences seems to
spring from such a scenario. Moreover, the energy scales of the
quintessence sector required to obtain the correct vacuum energy now
are minute as a consequence of the tiny value of the quintessence
field now. As a result, one must introduce a highly fine-tuned scale
in the quintessence sector which is as difficult to explain as the
smallness of the vacuum energy. On the other hand, one may hope to
preserve the SUGRA potential runaway shape by tuning the hidden sector
dynamics. In this case, we find that the local test of gravity (fifth
force, weak equivalence principle and proton to electron mass ratio
variation) are incompatible with the vev of the quintessence field
implied by the presence of an attractor which, again, is required if
we want insensitivity to the initial conditions. Hence it seems
difficult to build models of quintessence in supergravity where the
cosmological, gravitational and particle physics aspects are
compatible.  One of the plausible possibilities consists in getting
rid of the regularity of the K\"ahler potential for small values of
the quintessence field. This is what happens in no--scale models for
instance. The analysis of no-scale models leading to both
supersymmetry breaking and quintessence is left for future
work~\cite{BMnoscale}.


\clearpage \nocite{*}
\bibliography{quintmat}

\providecommand{\href}[2]{#2}\begingroup\raggedright\begin{thebibliography}{90}

\bibitem{LSS} Tegmark M {\it et al.}, Cosmological Parameters from
SDSS and WMAP, 2004 {\it Phys.~Rev.~D} {\bf 69} 103501
[astro-ph/0310723].

\bibitem{IA} Perlmutter S {\it et al.}, Measurements of Omega and
Lambda from 42 High-Redshift Supernovae, 1999 {\it Astrophys.~J.} {\bf
517} 565 [astro-ph/9812133]; Garnavich P M {\it et al.}, Constraints
on Cosmological from Hubble Space Telescope Observations of High-z
Supernovae, 1998 {\it Astrophys.~J.} {\bf 493} L53 [astro-ph/9710123];
Riess A G {\it et al.}, Observational Evidence from Supernovae for an
Accelerating Universe and Cosmological Constant, 1998 {\it Astron.~J.}
{\bf 116} 1009 [astro-ph/9805201].

\bibitem{CMB} Spergel D N {\it et al.}, Wilkinson Microwave Anisotropy
Probe (WMAP) Three Years Results: Implications for Cosmology
[astro-ph/0603449]; Fosalba P, Gaztanaga E and Castander F, Detection
of the ISW and SZ Effecst from CMB-Galaxy Correlation, 2003 {\it
Astrophys.~J.} {\bf 597} L89 [astro-ph/0307249]; Scranton R {\it et
al.}, Physical Evidence of Dark Energy [astro-ph/0307335]; Boughn S
and Crittenden R, A Correlation of the Cosmic Microwave Sky with Large
Scale Structure, 2004 {\it Nature (London)} {\bf 427} 45
[astro-ph/0305001].

\bibitem{kachru} Kachru S, Schulz M and Silverstein E, Self-Tuning of
Flat Domain Walls in 5d Gravity and String Theory, 2000 {\it
Phys.~Rev.~D} {\bf 62} 045021 [hep-th/001206]; Arkani-Hamed N,
Dimopoulos S, Kaloper N and Sundrum R, A Small Cosmological Constant
for a Large Extra Dimensions, 2000 {\it Phys.~Lett.} {\bf B480} 193
[hep-th/000197].

\bibitem{deffayet} Deffayet C, Dvali G and Gavadadze G, Accelerated
Universe from Gravity Leaking to Extra Dimension, 2002 {\it
Phys.~Rev.~D} {\bf 65} 044023 [hep-th/0105068].

\bibitem{Lalak} Forste S, Lalak Z, Lavignac S and Nilles H P, A
Comment on Self--Tuning and vanishing Cosmological Constant in the
Brane World, 2000 {\it Phys.~Lett.} {\bf B481} 360 [hep-th/0002164].


\bibitem{MSU} Martin J, Schimd C and Uzan J P, Testing for $w<-1$ in
  the Solar System, {\it Phys.~Rev.~Lett.} to be published
  [astro-ph/0510208].

\bibitem{RP} Ratra B and Peebles P J E, Cosmological Consequences of a
  Rolling Homogeneous Scalar Field, 1988 {\it Phys.~Rev.~D} {\bf 37}
  3406.

\bibitem{quint} Wetterich C, Cosmologies with Variable Newton's
``Constant'', 1988 {\it Nucl.~Phys.} {\bf B302} 668; Wetterich C, The
Cosmon Model for an Asymptotically Vanishing Time-Dependent
Cosmological Constant, 1995 {\it Astron.~Astrophys.}  {\bf 301} 321
[hep-th/9408025]; Ferreira P G and Joyce M, Cosmology with a
Primordial Scaling Field, 1998 {\it Phys.~Rev.~D} {\bf 58}, 023503
[astro-ph/9711102].

\bibitem{PB} Bin\'etruy P, Models of Dynamical Supersymmetry Breaking
  and Quintessence, 1998 {\it Phys.~Rev.~D} {\bf 60} 063502
  [hep-ph/9810553]; Bin\'etruy P, Cosmological Constant Versus
  Quintessence, 2000 {\it Int.~J.~Theor.~Phys.}  {\bf 39}, 1859
  [hep-ph/0005037].

\bibitem{Nilles} Nilles H P, Supersymmetry, Supergravity and Particle
  Physics, 1984 {\it Phys.~Rept.} {\bf 101} 1; Martin S P, A
  Supermmetry Primer, [hep-ph/9709356]; Aitchison I J R, Supersymmetry
  and the MSSM: An Elementary Introduction, Notes of Lectures for
  Graduate Students in particle Physics, Oxford 1004 \& 2005.

\bibitem{BM1} Brax P and Martin J, Quintessence and Supergravity, 1999
  {\it Phys.~Lett.} {\bf B468} 40 [astro-ph/9905040].

\bibitem{BM2} Brax P and Martin J, The Robustness of Quintessence,
  2000 {\it Phys.~Rev.~D} {\bf 61} 103502 [astro-ph/9912046].

\bibitem{BMR1} Brax P, Martin J and Riazuelo A, Exhaustive Study of
Cosmic Microwave Background Anisotropies in Quintessential Scenarios,
2000 {\it Phys.~Rev.~D} {\bf 62} 103505 [astro-ph/0005428].

\bibitem{BMR2} Brax P, Martin J and Riazuelo A, Quintessence with Two
  Energy Scales, 2001 {\it Phys.~Rev.~D} {\bf 64} 083505
  [hep-ph/0104240].

\bibitem{BMpart} Brax P and Martin J, Dark Energy and the MSSM,
[hep-th/0605228].

\bibitem{BMnoscale} Brax P and Martin J, No Scale Quintessence , in
preparation.

\bibitem{Ibanez} Ibanez, Locally Supersymmetric $SU(5)$ Grand
Unification, 1982 {\it Phys.~Lett.} {\bf B118} 73.

\bibitem{Brax} Brax P and Savoy C, Models with Inverse Sfermion Mass
 Hierarchy and Decoupling of the SUSY FCNC Effects, 2000 {\it JHEP}
 {\bf 0007} 048 [hep-ph/0004133].

\bibitem{Khoury} Das S, Corasiniti P S and Khoury J,
  Super-acceleration as Signature of Dark Sector Interaction,
  [astro-ph/0510628].

\bibitem{GR} Will C M, The Confrontation between General Relativity
  and Experiment, 2006 {\it Living. Rev. Rel.} {\bf 9} 2
  [gr-qc/0510072]; Fischbach E and Talmadge C, The Search for
  non-Newtonian Gravity, 1999 {\it Springer-Verlag, New-York};
  Bertotti B, Iess L and Tortora P, A Test of General Relativity Using
  Radio Links with the Cassini Spacecraft, 2003 {\it Nature} {\bf 425}
  374; Esposito-Farese G [gr-qc/0409081].

\bibitem{DP} Damour T and Polyakov A M, The String Dilaton and the
Least Coupling Principle, 1994 {\it Nucl. Phys. }{ \bf B423} 532
[hep-th/9401069].

\bibitem{damour} Damour T, Testing the Equivalence Principle: Why and
How?, 1996 {\it Class. Quantum Grav.} {\bf 13} A33 [gr-qc/9606080]. 

\bibitem{JP} Uzan J P, The Fundamental Constants and their Variations:
  Observational Status and Theoretical Motivations, 2003 {\it
  Rev. Mod. Phys.} {\bf 75} 403 [hep-ph/0205340].

\bibitem{Su} Su Y {\it et al}, New Tests of the Universality of Free
  Fall, 1994 {\it Phys.~Rev.~D} {\bf 50} 3614; Baessler {\it et al},
  Improved Test of the Equivalence Principle for Gravitational
  Self-Energy, 1999 {\it Phys. Rev. Lett.} {\bf 83} 3585; Adelberger E
  G, New Tests of Eisntein's Equivalence Principle and Newton's
  Inverse-Squared Law, 2001 {\it Class. Quantum Grav.} {\bf 18} 2397;
  Williams J G, Turyshev S G and Boggs D H, Progress in Lunar Ranging
  Tests of Relativistic Gravity, 2004 {\it Phys. Rev. Lett.} {\bf 93}
  261101 [gr-qc/0411113].

\bibitem{petit} Ivanchik A {\it et al}, Does the Photon-to-Electron
  Mass Ratio Vary in the Course of the Cosmological Evolution?  (2003)
  {\it Astrophys. Space. Sci.} {\bf 283} 583 (2003)
  [astro-ph/0210299].

\bibitem{reinhold} Reinhold E {\it et al}, Indication of a
  Cosmological Variation of the Proton-Electron Mass Ratio Based on
  Laboratory Measurement and Reanalysis of $\mbox{H}_2$ Spectra, 2006
  {\it Phys. Rev. Lett.}  {\bf 96} 151101.

\bibitem{cham} Brax P, van de Bruck C, Davis A C, Khoury J and Weltman
  A, Detecting Dark Energy in Orbit--the Cosmological Chameleon,
  (2004) {\it Phys. Rev. D } {\bf 70} 123518 [astro-ph/0408415].

\end{thebibliography}\endgroup
\bibliographystyle{plain}

\end{document}